\documentclass[journal]{IEEEtran}
\usepackage[normalem]{ulem}
\usepackage[noadjust]{cite}
\usepackage{bbm}
\usepackage{array}
\usepackage{dcolumn}
\usepackage{epsfig}
\usepackage[intlimits]{amsmath}
\usepackage{amsmath, amsfonts, yhmath, bm}
\usepackage{amssymb}
\usepackage{psfrag}
\usepackage{color,soul}
\usepackage[dvipsnames]{xcolor}
\usepackage[normalem]{ulem}
\usepackage{enumerate}
\usepackage{stackengine}
\usepackage[noadjust]{cite}
\usepackage{graphicx}
\usepackage{caption}
\usepackage{subcaption}
\usepackage[font=footnotesize]{subcaption}
\usepackage{multirow}
\usepackage{cite}
\usepackage[font=footnotesize]{caption}
\usepackage{etoolbox}
\usepackage{tcolorbox}
\usepackage{float}
\usepackage{dsfont}
\usepackage{tikz}
\usetikzlibrary{arrows}
\newcommand {\mymat}[1]  {{\mbox{\boldmath $#1$}}}

\newcommand*{\myfontb}{\fontfamily{lmr}\selectfont}

\DeclareMathAlphabet      {\mathbfit}{OML}{cmm}{b}{it}

\newcommand{\etal}{\textit{et al.}}

\newcommand {\Y} {\mymat{Y}}

\newcommand {\Rset} {\mathbb{R}}
\newcommand {\Cset} {\mathbb{C}}
\newcommand {\Zset} {\mathbb{Z}}
\newcommand {\Eset} {\mathbb{E}}
\newcommand {\Nset} {\mathbb{N}}

\newcommand {\tps} {\rm{T}}

\newcommand {\LS} {\text{\tiny LS}}
\newcommand {\sw} {\text{sw}}

\makeatletter
\newsavebox\myboxA
\newsavebox\myboxB
\newlength\mylenA

\newcommand*\mybar[2][0.75]{%
	\sbox{\myboxA}{$\m@th#2$}%
	\setbox\myboxB\null
	\ht\myboxB=\ht\myboxA%
	\dp\myboxB=\dp\myboxA%
	\wd\myboxB=#1\wd\myboxA
	\sbox\myboxB{$\m@th\overline{\copy\myboxB}$}
	\setlength\mylenA{\the\wd\myboxA}
	\addtolength\mylenA{-\the\wd\myboxB}%
	\ifdim\wd\myboxB<\wd\myboxA%
	\rlap{\hskip 0.5\mylenA\usebox\myboxB}{\usebox\myboxA}%
	\else
	\hskip -0.5\mylenA\rlap{\usebox\myboxA}{\hskip 0.5\mylenA\usebox\myboxB}%
	\fi}
\makeatother

\newtheorem{lem}{Lemma}

\newcommand {\DFT} {\textrm{DFT}}
\newtheorem{theorem}{Theorem}

\def\comment#1{}

\newcommand{\stkout}[1]{
	\color{red}\ifmmode\text{\sout{\ensuremath{#1}}}\else\sout{#1}\fi\color{black}}

\newcommand{\addra}{}
\newcommand{\delra}{\comment}
\newcommand{\delraeq}{\comment}

\newcommand{\addraRI}{}

\begin{document}

\title{``Self-Wiener" Filtering: Data-Driven Deconvolution of Deterministic Signals}

\author{Amir Weiss and Boaz Nadler
	
	\thanks{Amir Weiss is with the Department of Electrical Engineering and	Computer Science, Massachusetts Institute of Technology, Cambridge, MA 02139 USA (email: amirwei@mit.edu)}
	\thanks{Boaz Nadler is with the Department of Computer Science and Applied Mathematics, Faculty of Mathematics and Computer Science, Weizmann Institute of Science,	234 Herzl Street, Rehovot 7610001 Israel, (e-mail:
	boaz.nadler@weizmann.ac.il)}
}

\maketitle

\begin{abstract}
We consider the \delra{fundamental }problem of robust deconvolution, and particularly the recovery of an unknown deterministic signal convolved with a known filter and corrupted by additive noise. We present a novel, non-iterative data-driven approach. Specifically, our algorithm works in the frequency-domain, where it tries to mimic the optimal unrealizable\addra{ non-linear} Wiener-like filter as if the unknown deterministic signal were known. This leads to a threshold-type regularized estimator, where the threshold at each frequency is determined in a\delra{ fully} data-driven manner. We perform a theoretical analysis of our proposed estimator, and derive approximate formulas for its Mean Squared Error (MSE) at both low and high Signal-to-Noise Ratio (SNR) regimes. We show\delra{ analytically} that in the low SNR regime our method provides enhanced noise suppression, and in the high SNR regime it approaches the optimal unrealizable solution. Further, as we demonstrate in simulations, our solution is highly suitable for (approximately) bandlimited or frequency-domain sparse signals, and provides a significant gain of several dBs relative to other methods in the resulting MSE.
\end{abstract}

\vspace{-0.3cm}
\section{Introduction}\label{sec:intro}
\addra{Deconvolution is a ubiquitous task in signal processing}\delra{A ubiquitous task in signal processing is {\myfontb{\emph{deconvolution}}}} \cite{riad1986deconvolution}\delra{, namely inverting the action of some system (or channel) on an input, desired signal}. \addra{When the measured convolved signal is contaminated with noise, deconvolution algorithms must carefully balance between the bandwidth and the Signal-to-Noise Ratio (SNR)}\delra{A major difficulty arises when the measured convolved signal is contaminated with noise, wherein a careful balancing of bandwidth and Signal-to-Noise Ratio (SNR) is required} \cite{mendel2012maximum}. Robust deconvolution problems appear in a\delra{ wide} variety of applications, such as communication systems, controllers, image and video processing, audio signal processing and ground-penetrating radar data analysis, to name \delra{but }a few \cite{de2001concurrent,ljung1999system,tendero2013optimal,krishnan2009fast,subramaniam1996cepstrum,schmelzbach2015efficient,wu2011comparison}.

\delra{A common measure for the quality of the signal reconstructed from the noisy convolution measurements is the Mean Squared Error (MSE) between the unknown input signal and the deconvolved one}\addra{A common quality measure of deconvolution algorithms is the Mean Squared Error (MSE) between the unknown input signal and the deconvolved one}. When the unknown signal and the noise are both modeled as stochastic stationary processes with known Second-Order Statistics (SOSs), the optimal solution within the class of linear estimators is the celebrated Wiener filter \cite{wiener1964extrapolation,van2013detection}. \delra{Following Wiener's work, many other deconvolution methods have been proposed, most of them under some \textit{a priori} knowledge}\addra{Various authors extended Wiener's approach, often by incorporating additional assumptions} about the \delra{class of the unknown }input signals\addra{ or the noise}. For example, Berkhout \cite{berkhout1977least} derived the least-squares inverse filtering\delra{ for known noise SOSs,} assuming that the input signal is white, namely with a constant spectral level\delra{\addra{, for known noise SOSs}}.\addra{ Assuming the input signal and noise are stochastic, with a priori known upper and lower bounds on their spectra at each frequency, a minimax approach was proposed in \cite{eldar2005robust}.}\delra{In \cite{eldar2005robust}, Eldar proposed a minimax approach, assuming the input signal and noise are stochastic, with \delra{\textit{a-priori}}\addra{a priori} known upper and lower bounds on their spectra at each frequency.}

While the random signal model is suitable in some \delra{problems and scenarios}\addra{settings}, in others the input signal is better modeled as {\myfontb\emph{deterministic unknown}}. Several methods have been proposed for this signal model as well \cite{eldar2005robust,lucy1974iterative,patel2009shearlet,benhaddou2016deconvolution}. One example is the WaveD algorithm\addra{ \cite{johnstone2004wavelet},}\delra{ proposed by Johnstone \etal\ \cite{johnstone2004wavelet}, which is} based on \delra{hard }thresholding of \delra{a }wavelet \delra{expansion}\addra{coefficients} (see also \cite{cavalier2007wavelet}, Section II\delra{, for a concise description of WaveD}). While some of these algorithms offer considerable enhancement, their performance may be \delra{affected by}\addra{sensitive to their} tuning parameters, which either need to be set by the user, or require separate careful calibration. Another class of deconvolution algorithms are iterative \cite{bennia1990optimization,dhaene1994extended,pruksch1998positive,neveux2000constrained,welk2013algorithmic}. Some of these methods also require tuning parameters, such as the $\lambda$ parameter in \cite{bennia1990optimization}, controlling the balance between noise reduction and filtration errors\delra{, caused by the deviation from the ideal inverse filter when $\lambda$ is non-zero. For a detailed discussion on this topic, see \cite{bako2016improved}}.

\begin{figure}[t]
	\centering
	\tikzstyle{int}=[draw, fill=blue!20, minimum size=3em]
	\tikzstyle{sum}=[draw, circle, fill=blue!20]
	\tikzstyle{init} = [pin edge={to-,thin,black}]
	\begin{tikzpicture}[node distance=2cm,auto,>=latex']
	\node [int] (a) {$h[n]$};
	\node (b) [left of=a,node distance=2cm, coordinate] {a};
	\node [sum, pin={[init]above:$v[n]$}] (c) [right of=a] {$\Sigma$};
	\node [coordinate] (end) [right of=c, node distance=2cm]{};
	
	\path[->] (b) edge node {$x[n]$} (a);
	\path[->] (a) edge node {} (c);
	\draw[->] (c) edge node {$y[n]$} (end) ;
	\draw [color=gray,thick](-2.25,-0.6) rectangle (4.25,1.4);
	\node at (-2.25,1.05) [above=6mm, right=0mm] {\textsc{Generation}};
	\end{tikzpicture}
	\begin{tikzpicture}[node distance=2cm,auto,>=latex']
	\node [int] (a) {$g[n]$};
	\node (b) [left of=a,node distance=2cm, coordinate] {a};
	\path[->] (b) edge node {$y[n]$} (a);
	\node (c) [right of=a,node distance=2cm, coordinate] {a};
	\path[->] (a) edge node {$\widehat{x}[n]$} (c);
	\draw [color=gray,thick](-2.25,-0.6) rectangle (2.25,0.7);
	\node at (-2.25,0.45) [above=5mm, right=0mm] {\textsc{Reconstruction}};
	\end{tikzpicture}
	\caption{Block diagram of model \eqref{modeleq} (\textsc{``Generation"}), and the considered class of estimators, produced by filtering (\textsc{``Reconstruction"}). Note that in our framework, $g[n]$ may depend on the measurements $\{y[n]\}_{n=0}^{N-1}$.}
	\label{fig:block_diagram_model}\vspace{-0.6cm}
\end{figure}
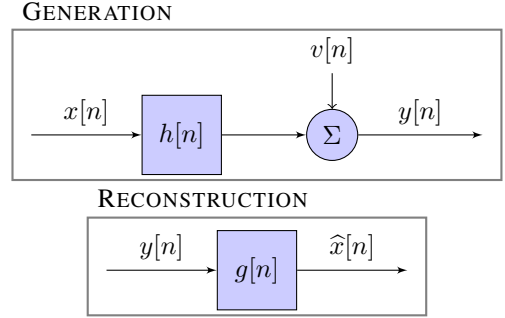

\delra{In this work, a}\addra{A}ssuming known\addra{ or estimated} SOS of the noise,\delra{ or an estimate thereof,} we propose a novel non-iterative, computationally simple, \delra{{\myfontb{\emph{fully}}}}\addra{fully} data-driven deconvolution approach for deterministic signals. The guiding principle of our approach, termed ``Self-Wiener" (SW) filtering, is an attempt to mimic the optimal Minimum MSE (MMSE) unrealizable\addra{\footnote{\addra{An ``unrealizable" solution is a solution which cannot be implemented in practice as it relies on additional (oracle) information that is not available.}}} Wiener-like filter, as if the unknown deterministic signal were known. \addra{This yields a thresholding-type method with no tuning parameters, where the threshold value is data dependent, bearing an intuitive interpretation. This is in contrast to other methods whose threshold is fixed (e.g., \cite{kalifa2003thresholding}).}\delra{This yields a\delra{ self-consistent} thresholding-type method with no tuning parameters\addra{,}\delra{. Our resulting threshold-type estimator is} reminiscent of other thresholding methods (e.g., \cite{kalifa2003thresholding})\addra{. However}, \delra{but }due to its data-driven nature, it automatically computes a data-dependent threshold value.}\delra{ Moreover, this threshold can be intuitively explained from an estimated SNR perspective.} We further present an analytical performance analysis of our proposed SW estimator, and derive approximate closed-form expressions for its \delra{predicted }MSE. Comparison in simulations to other approaches, some of which are fully data-driven as well, show that our method offers highly competitive performance, and attains an MSE lower by several dBs for various signals representative of those appearing in applications.

\addra{The rest of the paper is organized as follows. In Section \ref{sec:problemformulation} we formulate the problem. The optimal solution in the considered class of estimators illustrated in Fig.\ \ref{fig:block_diagram_model} is presented in Section \ref{sec:optimal_solution}. Our proposed estimator is derived in Section \ref{sec:swfilter}. In Section \ref{sec:MSEanalysis} we provide an analytical performance analysis. Empirical simulation results, corroborating our analytical derivation, are presented in Section \ref{sec:simulresults}, followed by concluding remarks in Section \ref{sec:conclusion}.}
\vspace{-0.4cm}
\section{Problem Formulation}\label{sec:problemformulation}
Let $\{y[n]\}_{n=1}^N$ be $N$ observations from the following classical discrete-time convolution model depicted in Fig.\ \ref{fig:block_diagram_model} (\textsc{``Generation"}), defined for all $n\in\Zset$,
\begin{equation}\label{modeleq}
y[n] = \sum_{k\in\Zset}{h[k]x[n-k]}+v[n]\in\Rset,\quad \forall n\in\Zset.
\end{equation}
Here, $h[n]$ is a known impulse response of a Linear Time-Invariant (LTI) system; $x[n]$ is an unknown deterministic signal; and $v[n]$ is a stationary, zero-mean additive noise with a \delra{positive }Power Spectral Density (PSD) function\delra{,} denoted by $S_{v}(\omega)$\addraRI{, but with an otherwise unknown distribution}. \addra{Note that the assumption that $v[n]$ is stationary does not imply that it is white or temporally uncorrelated. }We assume that the noise PSD $S_{v}(\omega)$ is either known or has been estimated \delra{\textit{a-priori}}\addra{a priori}, e.g., from realizations of pure noise, measured in a ``training period" \cite{stoica2005spectral}. We further assume that $x[n]$ is periodic or has finite support, and that $h[n]$ is \addra{exactly or may be well-approximated by }a Finite Impulse Response (FIR) filter, which \delra{is compactly supported }by definition\addra{ is compactly supported}\delra{, or at least may be well-approximated by an FIR filter}.
\vspace{-0.5cm}
\subsection{The Robust Deconvolution Problem}\label{subsec:robustdeconv}
The robust deconvolution problem \cite{walden1988robust} is to recover the signal values $\{x[n]\}_{n=1}^N$ based on the $N$ noisy measurements $\{y[n]\}_{n=0}^{N-1}$. Switching the roles of $x[n]$ and $h[n]$, yields the {\myfontb\emph{system identification}} problem \cite{aastrom1971system}. In that context, $x[n]$ is known and the problem is to estimate the unknown impulse response $h[n]$, namely to identify the system. Hence, while in this work we consider deconvolution, our proposed SW estimator is applicable to \delra{the }system identification \delra{problem }as well.

In the robust deconvolution context, the quality of an estimator $\widehat{x}[n]$ of $x[n]$ is often measured by its MSE,
\begin{equation}\label{costfunction}
\text{MSE}\left(x,\widehat{x}\right)\triangleq\Eset\left[\sum_{n=0}^{N-1}\left|x[n]-\widehat{x}[n]\right|^2\right],
\end{equation}
where the expectation is w.r.t.\ the noise $v[n]$ in the observations $y[n]$, the only random component in the problem. In this work, we focus on deconvolution methods of the following form, as depicted in Fig.\ \ref{fig:block_diagram_model} (\textsc{``Reconstruction"}),
\begin{equation}\label{solutionform}
\widehat{x}[n]=g[n]\otimes y[n],\; \forall n\in\{0,\ldots,N-1\},
\end{equation}
where $\otimes$ denotes circular convolution. The goal is to design a filter $g[n]$ that gives a low MSE. In contrast to the classical linear Wiener filter, we allow the filter $g[n]$ to depend on the observed signal $y[n]$, hence leading to a nonlinear estimator. To motivate our proposed filter, in Section \ref{sec:optimal_solution} we first study the optimal unrealizable filter of the form \eqref{solutionform}, which depends on the unknown signal $x[n]$. Next, in Section \ref{sec:swfilter} we derive a realizable data-driven estimator that is close to the optimal unrealizable solution at low and high SNRs. As demonstrated in Section \ref{sec:simulresults}, our estimator achieves MSEs that can be several dB lower than those obtained by other methods.
\vspace{-0.5cm}
\subsection{Equivalent Formulation in the Frequency Domain}\label{subsec:freq_formulation}
\addra{Due to our assumptions on the signal $x[n]$ and the filter $h[n]$, when the length $N$ of the observed output $y[n]$ is sufficiently large}\delra{Due to our assumptions regarding the input signal $x[n]$ and the FIR filter $h[n]$, given a finite number of samples $\{y[n]\}_{n=0}^{N-1}$, when $N$ is sufficiently large}, the linear convolution coincides or \delra{may be}\addra{is} well-approximated by circular convolution (neglecting boundary effects) \cite{sigprocoppen2009}. Hence, we consider the problem in the frequency domain. 

Recall that the unitary Discrete Fourier Transform (DFT) of a length-$N$ sequence $a[n]$ is defined as
\begin{equation}\label{DFTdef}
A[k] \triangleq \DFT\{a[n]\} \triangleq \frac{1}{\sqrt{N}}\sum_{n=0}^{N-1}{a[n]e^{-\jmath\frac{2\pi}{N}nk}}\in\Cset,
\end{equation}
for all $k\in\{0,\ldots,N-1\}$. Since circular convolution in the discrete-time domain is equivalent to multiplication in the (discrete-)frequency domain, applying the DFT to the sequence $\{y[n]\}_{n=0}^{N-1}$ \eqref{modeleq} gives
\begin{equation}\label{modeleqFreq}
Y[k] = H[k]X[k] + V[k],\; \forall k\in\{0,\ldots,N-1\},
\end{equation}
and \eqref{solutionform} becomes $\widehat{X}[k]=G[k]Y[k]$. Since the DFT is\delra{ a} unitary\delra{ transformation}, \delra{by Parseval's identity}
\begin{equation}\label{costfunctionFreq}
\text{MSE}\left(x,\widehat{x}\right)=\sum_{k=0}^{N-1}\Eset\left[\left|X[k]-\widehat{X}[k]\right|^2\right]\triangleq\sum_{k=0}^{N-1}\text{MSE}[k],
\end{equation}
where\delra{, with a slight abuse in notation,}\addra{ $\text{MSE}[k]$ is the MSE at the $k$-th frequency,}
\begin{equation}\label{kthMMSE}
\text{MSE}[k]\hspace{-0.06cm}=\hspace{-0.06cm}\Eset\left[\left|X[k]\hspace{-0.02cm}-\hspace{-0.02cm}\widehat{X}[k]\right|^2\right]\hspace{-0.07cm}=\hspace{-0.06cm}\Eset\left[\left|X[k]\hspace{-0.02cm}-\hspace{-0.02cm}G[k]Y[k]\right|^2\right]\addra{.}
\end{equation}
\delra{denotes the MSE at the $k$-th frequency component. }Obviously, {\myfontb{\emph{{separately}}}} minimizing each term $\text{MSE}[k]$ in the sum \eqref{costfunctionFreq}, minimizes the MSE \eqref{costfunction}. However,\addra{ as \delra{we }show\addra{n} in the next section,} when $x[n]$, or equivalently $X[k]$, is \delra{assumed }deterministic,\addra{ the optimal filter $G_{\text{opt}}[k]$ (given below in \eqref{optimalfilter}) depends on the unknown $X[k]$, and thus does not yield a realizable estimator}\delra{ the resulting optimal solution---$\widehat{X}_{\text{opt}}[k]$, to be introduced shortly in \eqref{optimalsolution} below---is not a realizable estimator, as we show next}.
\vspace{-0.6cm}
\section{The Optimal Deconvolution MMSE Solution}\label{sec:optimal_solution}
To motivate our proposed estimator, it is first instructive to present the\delra{ structure of the} optimal\delra{, yet not realizable} solution, which minimizes \eqref{costfunction} over all estimators of the form \eqref{solutionform}. For this, \delra{let us}\addra{we} begin by introducing\delra{ the following quantities}:
\begin{equation}\label{SNRout}
\begin{gathered}
\text{SNR}[k]\triangleq\frac{\left|X[k]\right|^2}{S_v[k]}, \quad \text{SNR}_{\text{out}}[k]\triangleq\frac{\left|H[k]X[k]\right|^2}{S_v[k]},
\end{gathered}
\end{equation}
where
\begin{equation*}
S_v[k]\triangleq S_v\left(\omega\right)\Big|_{\omega=\tfrac{2\pi k}{N}}=S_v\left(\tfrac{2\pi k}{N}\right)
\end{equation*}
is the noise PSD at the $k$-th frequency\addra{, assumed to be positive}. By definition, $\text{SNR}[k]$ and $\text{SNR}_{\text{out}}[k]$ are the SNRs at the $k$-th frequency of the \delra{noise-free} signal $x[n]$ and of its convolution with $h[n]$, at the output of the system, respectively.\delra{ Since $x[n]$ is unknown, both quantities are unknown.} Further, we define
\begin{equation}\label{LSestimate}
\addraRI{\widehat{X}_{\LS}[k]\triangleq\begin{cases}
Y[k]/H[k], & H[k]\neq0\\
0, & H[k]=0\\
	\end{cases},}
\delra{\widehat{X}_{\LS}[k]\triangleq\delra{\frac{Y[k]}{H[k]}}\addra{Y[k]/H[k]}, \quad \forall k\in\{0,\ldots,N-1\},}
\end{equation}
which is the na\"ive Least-Squares (LS) estimator of $X[k]$, obtained by filtering the noisy measurements using the inverse of the known filter $H[k]$\delra{, assuming $H[k]\neq0$}.

\addraRI{ Clearly, as seen from \eqref{modeleqFreq}, at frequencies where $H[k]=0$, the information about $X[k]$ is completely lost, and accordingly $Y[k]$ contains only the noise component $V[k]$. Since the filter $H[k]$ is assumed to be known, the optimal \emph{realizable} filter at such frequencies is $G_{\text{opt}}[k]=0$, eliminating the noise (as shown below in \eqref{optimalfilter}). Therefore, to facilitate the derivations throughout the paper, we assume hereafter without loss of generality that $H[k]\neq0$ for all $k\in\{1,\ldots,K\}$.}

The optimal MMSE filter $G_{\text{opt}}[k]$ may be found by differentiating \eqref{kthMMSE} w.r.t.\ $G^*[k]$ and equating to zero\addraRI{ \cite{brown2009complex}}. This gives
\begin{equation}\label{partialderivative}
-\left|X[k]\right|^2H^*[k]+G[k]\Eset\left[\left|X[k]H[k]+V[k]\right|^2\right]=0,
\end{equation}
where we have used $\Eset\left[V^*[k]\right]=0$. Since the noise $v[n]$ is stationary, for a sufficiently large $N$ (see e.g.\ \cite{gray2006toeplitz}),
\begin{equation*}
\Eset\left[\left|V[k]\right|^2\right]\underset{N\gg1}{\approx} S_v[k].
\end{equation*}
Thus, after simplifying and arranging the terms in \eqref{partialderivative},
the optimal deconvolving filter at the $k$-th frequency is
\begin{equation}\label{optimalfilter}
G_{\text{opt}}[k]\triangleq\frac{H^*[k]}{\left|H[k]\right|^2+\frac{S_v[k]}{\left|X[k]\right|^2}}=\frac{H^*[k]}{\left|H[k]\right|^2+\frac{1}{\text{SNR}[k]}}.
\end{equation}
Accordingly, the corresponding optimal solution is given by
\begin{equation}\label{optimalsolution}
\widehat{X}_{\text{opt}}[k]=Y[k]G_{\text{opt}}[k]=\widehat{X}_{\LS}[k]\cdot\frac{1}{1+\frac{1}{\text{SNR}_{\text{out}}[k]}}.
\end{equation}
The resulting MMSE at the $k$-th frequency in the class of estimators of the form \eqref{solutionform} is thus
\begin{gather}\label{MMSEexpression}
\begin{aligned}
\text{MMSE}[k]&\triangleq\Eset\left[\left|X[k]-\widehat{X}_{\text{opt}}[k]\right|^2\right]\\
&=\frac{\left|X[k]\right|^2}{1+\text{SNR}_{\text{out}}[k]}=\sigma_{\text{eff}}^2[k]\cdot\frac{1}{1+\tfrac{1}{\text{SNR}_{\text{out}}[k]}},
\end{aligned}
\end{gather}
where $\sigma_{\text{eff}}^2[k]$ is the effective noise level at the system output\addra{'s} $k$-th frequency, defined for all $k\in\{0,\ldots,N-1\}$ as
\begin{equation}\label{effnoise}
\sigma_{\text{eff}}^2[k]\triangleq\delra{\frac{S_v[k]}{|H[k]|^2}}\addra{S_v[k]/|H[k]|^2}.
\end{equation}

Note that the MMSE \eqref{MMSEexpression} is equal to the effective noise level at the output of the system multiplied by a \addra{regularization} term, which also depends on the unknown signal $x[n]$. However, the corresponding optimal filter does not yield a realizable estimator, as it depends on $\text{SNR}[k]$, which in turn depends on the unknown signal $x[n]$, as seen by \eqref{optimalfilter} and \eqref{SNRout}, respectively.
\delra{As seen from \eqref{optimalfilter} and \eqref{SNRout}, the optimal deconvolution filter does not yield a realizable estimator, since it depends on $\text{SNR}[k]$, which in turn depends on the unknown signal $x[n]$. Accordingly, the MMSE \eqref{MMSEexpression} is equal to the effective noise level at the output of the system multiplied by a \delra{regularizing}\addra{regularization} term, which also depends on the unknown signal $x[n]$.}\delra{ Further, observe that when $|X[k]|^2\to\infty$ while $\sigma_{\text{eff}}^2[k]$ is fixed, and thus $\text{SNR}_{\text{out}}[k]\to\infty$, the MMSE \eqref{MMSEexpression} reads}
\delraeq{\begin{equation*}\label{MMSElimitXinf}
\delra{\text{MMSE}[k]\xrightarrow[\sigma_{\text{eff}}^2\text{ fixed}]{\text{SNR}_{\text{out}}[k]\to\infty}\sigma_{\text{eff}}^2[k]\,.}
\end{equation*}}
\delra{Hence infinite output SNR does not guarantee perfect reconstruction at the output of the system even for the optimal unrealizable solution. However, when $\sigma_{\text{eff}}^2[k]\to0$ while $X[k]$ is fixed, which also implies $\text{SNR}_{\text{out}}[k]\to\infty$, the MMSE \eqref{MMSEexpression} converges to zero, which corresponds to perfect reconstruction.\addra{ We conclude that this (only apparent) discrepancy is due to the complete lack of knowledge of the desired signal $X[k]$.}}

It is interesting to note the resemblance of $G_{\text{opt}}[k]$ to the Wiener filter. For an input signal $x[n]$ modeled as stationary stochastic process with a known PSD, the Wiener filter has the form \eqref{optimalfilter}, but with $\left|X[k]\right|^2$ replaced by the PSD of $x[n]$.
\vspace{-0.3cm}
\section{``Self-Wiener" Filtering}\label{sec:swfilter}
\addraRI{\subsection{Main Results}\label{subsec:mainresults}}
The structure of the optimal solution \eqref{optimalsolution} motivates the following iterative approach: Start from an initial estimate of $X[k]$, and use it to estimate the quantity $\text{SNR}_{\text{out}}[k]$ defined in \eqref{SNRout}. Then, plug this into \eqref{optimalsolution} to obtain an improved estimate of $X[k]$. This principle leads to the following iterative procedure,
\begin{equation}\label{iterativeest}
\widehat{X}^{(t+1)}_{\sw}[k]=\widehat{X}_{\LS}[k]\cdot\frac{1}{1+\frac{\sigma_{\text{eff}}^2[k]}{\left|\widehat{X}^{(t)}_{\sw}[k]\right|^2}},\quad \forall t\in\Nset_0,
\end{equation}
starting from some $\widehat{X}^{(0)}_{\sw}[k]$.\addra{ This approach is reminiscent of the one taken in \cite{hiller1990iterative}, wherein the signal of interest is a two-dimensional image, assumed to be random, with an unknown auto-covariance matrix.} The following theorem shows that, when initialized with the LS estimator \eqref{LSestimate}, these iterations converge to a limit with a simple explicit form. The result described in \eqref{sw_est} below is our proposed estimator.
\begin{theorem}\label{theorem1}
\textit{Let $\widehat{X}^{(0)}_{\emph{\sw}}[k]=\widehat{X}_{\emph{\LS}}[k]$. Then, at each frequency $k$, the iterations \eqref{iterativeest} converge to a solution $\widehat{X}_{\textrm{\emph{\sw}}}[k]$, which satisfies
\begin{equation}\label{implicitWiener}
\widehat{X}_{\textrm{\emph{\sw}}}[k]=\widehat{X}_{\emph{\LS}}[k]\cdot \frac{1}{1+\frac{\sigma_{\text{\emph{eff}}}^2[k]}{\left|\widehat{X}_{\emph{\sw}}[k]\right|^2}}.
\end{equation}
The solution of \eqref{implicitWiener} is the following thresholding operator}
\tcbset{colframe=gray!95!blue,size=small,width=0.49\textwidth,arc=2.1mm,outer arc=1mm}
\begin{tcolorbox}[upperbox=visible,colback=white,boxsep=1.5pt,bottom=1.5pt]
\vspace{-0.15cm}
\begin{equation}\label{sw_est}
	\widehat{X}_{{\textrm{\sw}}}[k]\triangleq\widehat{X}_{\LS}[k]\cdot\begin{cases}
	\frac{2\left|Z[k]\right|^{-2}}{1-\sqrt{1-4\left|Z[k]\right|^{-2}}},& \left|Z[k]\right|>2\\
	0,& \left|Z[k]\right|<2
	\end{cases},
\end{equation}
\end{tcolorbox}
\noindent \textit{where}
\begin{equation}\label{Zdef}
Z[k]\triangleq\delra{\frac{Y[k]}{\sqrt{S_v[k]}}}\addra{Y[k]/\sqrt{S_v[k]}}, \; \forall k\in\{0,\ldots,N-1\}.
\end{equation}
\end{theorem}
In the proof of Theorem \ref{theorem1}, we use the following result from stability theory (e.g., \cite{Holmes:2006}):
\begin{theorem}\label{theorem2}
	\textit{Let $\gamma_*$ be a fixed point of a continuously differentiable function $f:\Rset\rightarrow\Rset$\addraRI{, namely $f(\gamma_*)=\gamma_*$}. Then $\gamma_*$ is stable if $\left|f'(\gamma_*)\right|<1$, and unstable if $\left|f'(\gamma_*)\right|>1$.}
\end{theorem}
\noindent\textit{Proof of Theorem \ref{theorem1}:} First, notice that the phase of the estimator \eqref{iterativeest} remains constant throughout the iterative process. Further, note that the phase of the optimal MMSE solution \eqref{optimalsolution} is equal to the phase of the LS estimator \eqref{LSestimate}. Hence, we focus on the convergence of the proposed estimator's magnitude. Further\delra{more}, for ease of notation, let us define the following three quantities, omitting \addra{the dependence on }$k$ for brevity:\delra{ $\gamma_t\triangleq\frac{1}{\left|\widehat{X}^{(t)}_{\sw}[k]\right|}$, the reciprocal magnitude of $\widehat{X}^{(t)}_{\sw}[k]$, and} $\alpha\triangleq\left|\widehat{X}_{\LS}[k]\right|^{-1}$, $\beta^2\triangleq\sigma_{\text{eff}}^2[k]$, which are constants throughout the iterative process\delra{, independent of the iteration index, $t$}\addra{, and $\gamma_t\triangleq\left|\widehat{X}^{(t)}_{\sw}[k]\right|^{-1}$}.
\delra{With these notations, t}\addra{T}he iterations \eqref{iterativeest} \addra{now }take the form
\begin{equation}\label{iterativeest2}
\gamma_{t+1}=\alpha\cdot\left(1+\beta^2\cdot\gamma^2_t\right)\triangleq f(\gamma_t)\in\Rset^+,\quad \forall t\in\Nset_0.
\end{equation}
Note that convergence of $\gamma_{t}$ is equivalent to convergence of $\left|\widehat{X}^{(t)}_{\sw}[k]\right|$. The convergence of\addra{ \eqref{iterativeest2}}\delra{ this recursive formula may}\addra{ can} be analyzed using \delra{principles from }stability theory. First, due to the randomness of $\alpha$, the quadratic equation $f(\gamma)=\gamma$ has either two solutions or none almost surely (i.e., it has one solution with zero probability). Thus, when $\alpha^2\beta^2>1/4$, there is no fixed point, and since $\gamma_{t}\geq0$, it follows that $f'(\gamma_{t})>0$ for all $t\in\Nset_0$, and $\gamma_{t}$ diverges. When $\alpha^2\beta^2<1/4$, there are two fixed points, denoted as $\gamma_*$ and $\widetilde{\gamma}_*$. It is easy to check that for one of these fixed points $|f'(\gamma_*)|<1$, whereas for the other $|f'(\widetilde{\gamma}_*)|>1$. Hence, by Theorem \ref{theorem2}, only the fixed point $\gamma_*$ is stable. Therefore, \delra{with only one \delra{of them }stable. \delra{Applying Theorem \ref{theorem2} to our transformed iterative procedure \eqref{iterativeest2}, one easily obtains that starting with $\gamma_0=\alpha$,}\addra{By Theorem \ref{theorem2},} starting from $\gamma_0=\alpha$,}
\begin{equation}\label{convergenceresult}
\gamma_{t}\xrightarrow[t\rightarrow\infty]{}\begin{cases}
\frac{1-\sqrt{1-4\alpha^2\beta^2}}{2\alpha\beta^2}\triangleq\gamma_*,& \alpha^2\beta^2<1/4\\
\infty,& \alpha^2\beta^2>1/4
\end{cases}.
\end{equation}
Since $\left(\alpha\beta\right)^{-2}=\left|Z[k]\right|^2$, inverting \eqref{convergenceresult} and multiplying by $\widehat{X}_{\LS}[k]/\left|\widehat{X}_{\LS}[k]\right|$ yields \eqref{sw_est}.\hfill $\blacksquare$\vspace{0.1cm}

\delra{Note that since $f(\gamma)$ is quadratic, when $\alpha^2\beta^2<1/4$, the iterative process in fact converges to $\gamma_*$ from any initial value $\gamma_0\in\left[0,\gamma_*\right]$. Specifically, with $\gamma_0=0$, rather than $\gamma_0=\alpha$ as suggested in the theorem, we \delra{again }obtain $\gamma_1=\alpha$, corresponding to the initial solution $\widehat{X}^{(1)}_{\sw}[k]=\widehat{X}_{\LS}[k]$.}

An equivalent, yet different \delra{enlightening}\addra{instructive} expression for the SW estimator \eqref{sw_est} is given as follows. Focusing on the case $|Z[k]|>2$, by multiplying the numerator and denominator in \eqref{sw_est} by $1+\sqrt{1-4|Z[k]|^{-2}}$, we obtain
\begin{equation}\label{sw_shrinkage_form}
\widehat{X}_{\textrm{\sw}}[k]=\widehat{X}_{\LS}[k]\cdot\frac12\left(1+\sqrt{1-4|Z[k]|^{-2}}\,\right).
\end{equation}
The form \eqref{sw_shrinkage_form} illustrates the {\myfontb\emph{shrinkage}} of the SW filter w.r.t.\ the nai\"ve LS estimator, with the shrinkage factor depending on the observed value $Z[k]$. At a high output SNR, with high probability $|Z[k]|\gg1$, and there is almost no shrinkage. In contrast, if $|Z[k]|\to2$ from above, the shrinkage tends to $1/2$. \delra{In the complementary case}\addra{For} $|Z[k]|<2$, the shrinkage factor is zero. \delra{Interestingly, this behavior of the proposed SW estimator is reminiscent of thresholding methods in high\addra{-}dimensional statistics}\addra{Hence, the SW estimator bears similarity to thresholding methods in statistics}, \delra{in particular}\addra{particularly} in the presence of sparsity (e.g., \cite{donoho1995adapting}).\addraRI{ These thresholding methods are highly efficient for sparse signals, some of which are even optimal under certain conditions \cite{donoho1995noising,tibshirani1996regression,fan2001variable}.} We thus expect our estimator to be superior to other, non-threshold-type estimators for signals with low energy in certain frequency components, such as bandlimited or sparse frequency-domain signals \delra{(e.g., }\cite{tropp2009beyond,dikmese2016sparse}\delra{)}. This will be illustrated via simulations in Section \ref{sec:simulresults}.
\vspace{-0.4cm}
\subsection{Comparison to the Optimal Unrealizable Solution}\label{subsec:comparison2optimal}
Next, let us compare the SW estimator to the unrealizable optimal solution \eqref{optimalsolution}. The latter is the LS estimator multiplied by a shrinkage factor, which depends on the unknown $X[k]$. Similarly, the proposed estimator\delra{, which is the solution of \eqref{implicitWiener},}\addra{ \eqref{sw_est}} is also the LS estimator multiplied by a \delra{regularization}\addra{shrinkage} factor with the same structure as the optimal one. However, since $X[k]$ is unknown, our\delra{ self-consistent} estimator ``uses itself" to construct the \delra{regularization term}\addra{resulting shrinkage, see \eqref{implicitWiener}}. Hence the name of the proposed method---``Self-Wiener" filtering. This intuition can be rigorously justified in the high SNR regime, as shown next. 
\vspace{-0.4cm}
\subsection{The \addra{``}Self-Wiener\addra{"} Estimator in the High SNR Regime}\label{subsec:highSNR}
\delra{Let us present another interpretation which sheds light from a different angle on the successful mode of operation of our proposed estimator \eqref{sw_est}. Specifically, we \delra{now }show that \delra{our estimator}\addra{in the high SNR regime \eqref{sw_est}} approximately coincides with the optimal solution \eqref{optimalsolution}\delra{ in the high SNR regime}.}

Recall that since $Y[k]=H[k]X[k]+V[k]$, then
\begin{equation}\label{defeta}
Z[k]=\tfrac{H[k]X[k]}{\sqrt{S_v[k]}}+\tfrac{V[k]}{\sqrt{S_v[k]}}\triangleq\eta[k]+\widetilde{V}[k]\delra{.}\addra{,}
\end{equation}
\delra{Hence}\addra{and} $\text{SNR}_{\text{out}}[k]\hspace{-0.05cm}=\hspace{-0.05cm}\left|\eta[k]\right|^2$\delra{, which}\addra{. We thus} naturally\delra{ leads to the} \delra{definition}\addra{define}
\begin{equation}\label{ZasSNRout}
\widehat{\text{SNR}}_{\text{out}}[k]\triangleq\left|Z[k]\right|^2=\left|\eta[k]+\widetilde{V}[k]\right|^2.
\end{equation}
At a high output SNR, where $\left|\eta[k]\right|\gg1$, with high probability $\left|Z[k]\right|\gg2$, and thus $4\left|Z[k]\right|^{-2}\ll1$. Using the second-order Taylor expansion $\sqrt{1-x}\approx1-\tfrac{x}{2}-\tfrac{x^2}{8}$ (valid for any $|x|\ll1$),
\begin{equation}\label{Taylorapprox}
\frac{2\left|Z[k]\right|^{-2}}{1-\sqrt{1-4\left|Z[k]\right|^{-2}}}\approx\frac{1}{1+\left|Z[k]\right|^{-2}}.
\end{equation}
Combining \eqref{ZasSNRout} and \eqref{Taylorapprox} yields that at $\text{SNR}_{\text{out}}[k]\gg1$,
\begin{equation}\label{highSNRestimate}
\widehat{X}_{\sw}[k]\approx\widehat{X}_{\LS}[k]\cdot\delra{\frac{1}{1+\frac{1}{\widehat{\text{SNR}}_{\text{out}}[k]}}}\addra{\left(1+1/\widehat{\text{SNR}}_{\text{out}}[k]\right)^{-1}}.
\end{equation}
Evidently, \eqref{highSNRestimate} has the same structure as the optimal solution \eqref{optimalsolution}, but uses the estimated output SNR rather than the true output SNR.\delra{ Hence, our proposed estimator is a nearly optimal fully data-driven approximation of \eqref{optimalsolution} when $X[k]$ is unknown.}\addra{ Additionally, \eqref{ZasSNRout} sheds light on the interpretation of $|Z[k]|$, and particularly on the condition $|Z[k]|>2$. If the estimated output SNR is sufficiently high, $\widehat{X}_{\sw}[k]$ ``wears his best Wiener disguise", which improves as the true SNR increases. Otherwise, namely at low estimated output SNR, it prefers zeroing the output.}


\delra{Our approach is also applicable when the noise spectrum $S_v[k]$ is not known a priori. Specifically, let $\widehat{S}_v[k]$ be an estimate of the noise spectrum, obtained either from separate noise-only realizations, or directly from the measured noisy convoluted data (as in \cite{donoho1995wavelet} for white noise). Then, we define $\widehat{Z}[k]\triangleq Y[k]/\sqrt{\widehat{S}_v[k]}$, and replace $Z[k]$ with $\widehat{Z}[k]$ in \eqref{sw_est}.}\addraRI{Our approach is also applicable when the noise spectrum $S_v[k]$ is not known, but we have an estimate of it, $\widehat{S}_v[k]$. This estimate may be obtained either from separate noise-only realizations, or in the case of white noise, directly from the measured noisy convoluted signal, e.g., as in \cite{donoho1995wavelet,john2021adaptive}.}


\delra{We now turn to a statistical performance analysis of the proposed estimator, which allows to accurately assess its performance in terms of its MSE.}
\vspace{-0.2cm}
\section{MSE Analysis of the Proposed Estimator}\label{sec:MSEanalysis}
\addra{We now present a statistical performance analysis  of the SW estimator, and derive approximate analytic formulas for its MSE. }\delra{In this section we derive closed-form approximate expressions for the MSE of our proposed SW estimator \eqref{sw_est}. }Since our estimator operates in the frequency domain,
\begin{equation*}\label{SW_est_MSE}
\text{MSE}\left(x,\widehat{x}_{\sw}\right)=\sum_{k=0}^{N-1}\Eset\left[\left|X[k]-\widehat{X}_{\sw}[k]\right|^2\right]\triangleq\sum_{k=0}^{N-1}\text{MSE}_{\sw}[k]\delra{,}\addra{.}
\end{equation*}
\addra{Hence, we may separately analyze the MSE at the $k$-th frequency, $\text{MSE}_{\sw}[k]$. Since the signal $x[n]$ is considered deterministic, the expectation in $\text{MSE}_{\sw}[k]$ is only over the random noise $V[k]$, which in turn is a function of $\{v[n]\}_{n=0}^{N-1}$ of \eqref{modeleq}.} We also denote $p[k]\triangleq\Pr\left(\left|Z[k]\right|>2\right)$, \delra{which is }where $|Z[k]|$ is given in \eqref{Zdef}. \delra{ Thus, we shall focus on the MSE at the $k$-th frequency component, $\text{MSE}_{\sw}[k]$.} Using \eqref{sw_est} and the law of total expectation,
\begin{equation}\label{MSElawoftotalexpectation}
\begin{aligned}
\hspace{-0.1cm}\text{MSE}_{\sw}[k]=\, p[k]\cdot\,&\Eset\left[\left.\left|X[k]-\widehat{X}_{\sw}[k]\right|^2\right|\left|Z[k]\right|>2\right]+\\
(1-p[k])\cdot\,&\Eset\left[\left.\left|X[k]-\underbrace{\widehat{X}_{\sw}[k]}_{=0}\right|^2\right|\left|Z[k]\right|\leq2\right].
\end{aligned}
\end{equation}
Inserting\addra{ into \eqref{MSElawoftotalexpectation}} the following relation 
\begin{equation*}\label{openabssqr}
\left|X[k]-\widehat{X}_{\sw}[k]\right|^2\hspace{-0.05cm}=\hspace{-0.05cm}\left|X[k]\right|^2+\left|\widehat{X}_{\sw}[k]\right|^2-2\Re\left\{\widehat{X}_{\sw}[k]X^*[k]\right\}\addra{,}
\end{equation*}
\delra{into \eqref{MSElawoftotalexpectation}, }we obtain after simplification
\begin{equation}\label{MSE_of_SW}
\begin{aligned}
\hspace{-0.19cm}\text{MSE}_{\sw}[k]\hspace{-0.05cm}&=\hspace{-0.05cm}\left|X[k]\right|^2\hspace{-0.05cm}+\hspace{-0.05cm}p[k]\hspace{-0.05cm}\cdot\hspace{-0.05cm}\Eset\left[\left.\left|\widehat{X}_{\sw}[k]\right|^2\right|\left|Z[k]\right|>2\right]\\
&-2p[k]\hspace{-0.05cm}\cdot\hspace{-0.05cm}\Re\left\{X^*[k]\cdot\Eset\left[\left.\widehat{X}_{\sw}[k]\right|\left|Z[k]\right|>2\right]\right\}.
\end{aligned}
\end{equation}

From this point on, we assume that the noise $V[k]$ is a Complex Normal (CN) Random Variable (RV). Thus, \delra{it follows that} $Z[k]\hspace{-0.05cm}\sim\hspace{-0.05cm}\mathcal{CN}(\eta[k],1)$, where $\eta[k]$ is defined in \eqref{defeta}. Indeed, this holds when the time-domain additive noise $v[n]$ is Gaussian. As seen from the DFT definition \eqref{DFTdef}, for a sufficiently large $N$ this is also approximately true for other noise distributions (under mild conditions) due to the Central Limit Theorem (CLT) \cite{heyde1974central}. In addition, for the sake of brevity, throughout this section we omit the frequency index $k$ in what follows. However, we emphasize that the following analysis is {\myfontb\emph{per frequency}}, and the overall performance depends on the sum of all the MSEs at all frequencies. \delra{Additionally, h}\addra{H}ere we analyze only the complex-valued frequency bins\delra{, corresponding to all the indices except} \delra{for }$k\delra{=}\addra{\neq}0,N/2$. The \delra{la\addra{t}ter, which correspond to }real-valued DFT components, \addra{corresponding to $k=0,N/2$, }are \delra{addressed}\addra{analyzed} in Appendix \ref{AppArealDFT}\delra{ (in a similar manner)}.

In order to obtain \delra{general }closed-form expressions of the MSE \eqref{MSE_of_SW}, one must compute the conditional expectations
\begin{align}
&\Eset\left[\left.\widehat{X}_{\sw}\right|\left|Z\right|>2\right]\hspace{-0.05cm}=\hspace{-0.05cm}\frac{\sqrt{S_v}}{H}\Eset\left[\left.\tfrac{2(Z^*)^{-1}}{1-\sqrt{1-4\left|Z\right|^{-2}}}\right|\left|Z\right|>2\right]\label{condexpc1},\\
&\Eset\left[\left.\left|\widehat{X}_{\sw}\right|^2\right|\left|Z\right|>2\right]\hspace{-0.05cm}=\hspace{-0.05cm}\frac{S_v}{|H|^2}\Eset\left[\left.\tfrac{4|Z|^{-2}}{\left(1-\sqrt{1-4\left|Z\right|^{-2}}\right)^2}\right|\left|Z\right|>2\right]\hspace{-0.075cm},\label{condexpc2}
\end{align}
as well as an expression for the probability $p$. In the next subsections we derive relatively simple approximations to these expectations at both low and high SNR regimes, leading to insightful approximate expressions of the SW estimator's MSE \eqref{MSE_of_SW}. We begin with the probability $p$.

\begin{figure}[t]
	\centering
	\includegraphics[width=0.38\textwidth]{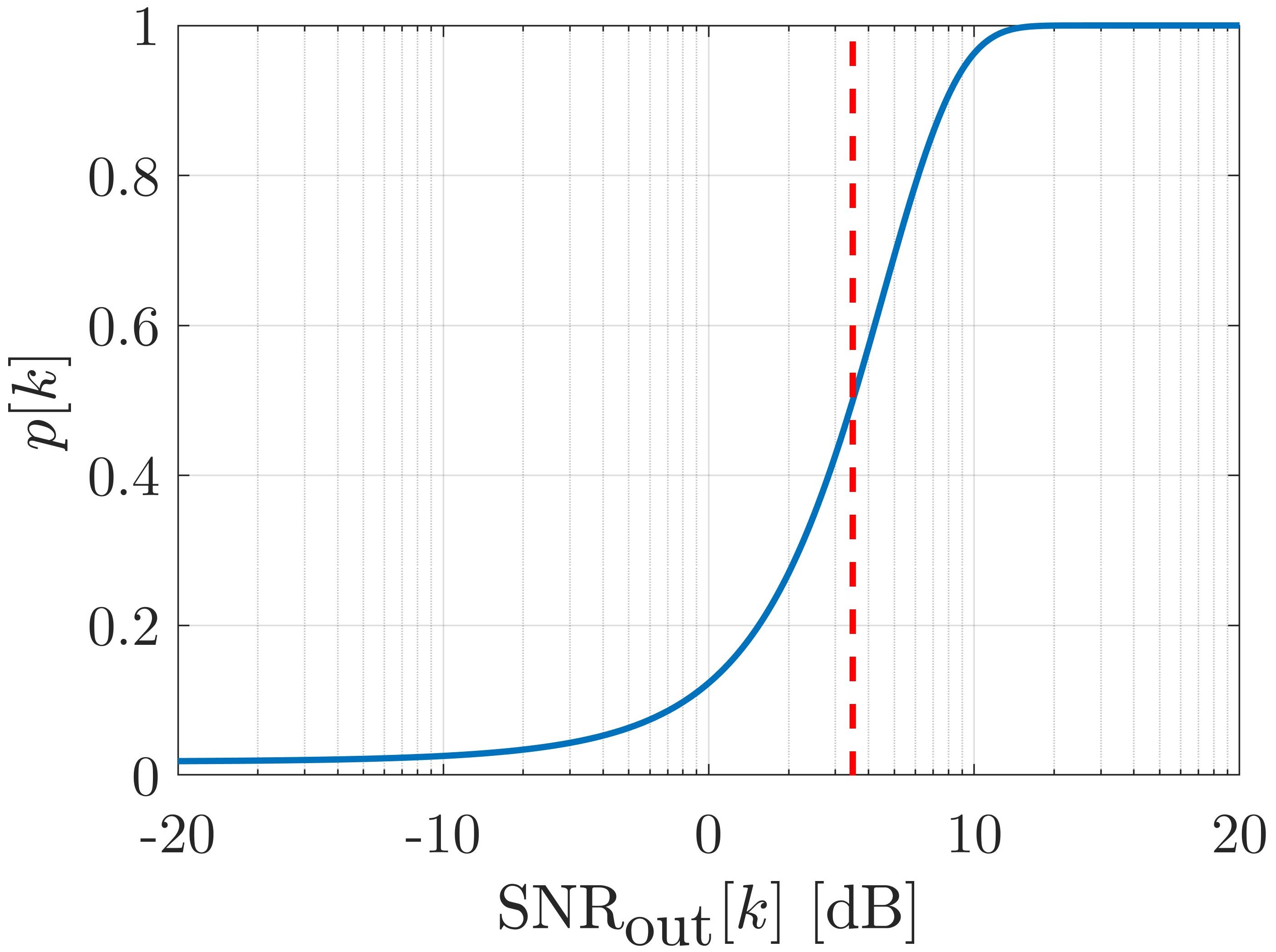}
	\caption{Probability of the $k$-th frequency to be above the threshold vs.\ the output SNR \eqref{SNRout}. The dashed red line \delra{marks}\addraRI{crosses} the point where $p[k]=0.5$\comment{, at $\text{SNR}_{\text{out}}\approx3.488$}.} 
	\label{fig:p_vs_SNRout}\vspace{-0.45cm}
\end{figure}

Using the notations $Z_{\text{r}}\triangleq\Re\{Z\}$ and $Z_{\text{i}}\triangleq\Im\{Z\}$, such that
\begin{equation*}\label{realandimagstat}
Z_{\text{r}}\sim\mathcal{N}\left(\Re\{\eta\},\tfrac{1}{2}\right),\; Z_{\text{i}}\sim\mathcal{N}\left(\Im\{\eta\},\tfrac{1}{2}\right),
\end{equation*}
we have
\begin{align*}\label{compofprob}
p&=\Pr\left(\left|Z\right|>2\right)=\Pr\left(\left|Z\right|^2>4\right)\\
&=\Pr\left(\left(\sqrt{2}Z_{\text{r}}\right)^2+\left(\sqrt{2}Z_{\text{i}}\right)^2>8\right)\triangleq \Pr\left(\Xi>8\right),
\end{align*}
where $\Xi$ is by definition a non-central chi-square RV with two degrees of freedom and a non-centrality parameter $2\left|\eta\right|^2$. Thus,\addra{ recalling that $\text{SNR}_{\text{out}}=|\eta|^2$,}
\begin{equation}\label{finalexpprob}
p=Q_{1}\left(\sqrt{2}\cdot\left|\eta\right|,2\sqrt{2}\right)=Q_{1}\left(\sqrt{2\cdot\text{SNR}_{\text{out}}},2\sqrt{2}\right),
\end{equation}
where \addra{$Q_{M}(a,b)$ is the Marcum Q-function \cite{nuttall1975some}.}\delra{we have used $\text{SNR}_{\text{out}}=|\eta|^2$, and}
\delra{is the Marcum Q-function \cite{nuttall1975some}, \delra{where}\addra{and} $I_{M-1}(\cdot)$ is the modified Bessel function of order $M-1$.} It is easy to verify that,
\begin{equation}\label{marcumqhighsNR}
p=Q_{1}\left(\sqrt{2\cdot\text{SNR}_{\text{out}}},2\sqrt{2}\right)\xrightarrow[]{\text{SNR}_{\text{out}}{}\rightarrow\infty}1,
\end{equation}
as illustrated in Fig.\ \ref{fig:p_vs_SNRout}. Hence, at frequencies with high output SNR, \eqref{MSElawoftotalexpectation} becomes
\begin{equation}\label{MSElimithighSNR}
\text{MSE}_{\sw}\xrightarrow[]{\text{SNR}_{\text{out}}{}\rightarrow\infty}\Eset\left[\left.\left|X-\widehat{X}_{\sw}\right|^2\right|\left|Z\right|>2\right],
\end{equation}
where $\widehat{X}_{\sw}$ in \eqref{MSElimithighSNR} is given by \eqref{highSNRestimate}. Therefore, in compliance with the interpretation given in Subsection \ref{subsec:highSNR}, the MSE of the SW estimator at high output SNR is nearly the MMSE \eqref{MMSEexpression} of the optimal solution \eqref{optimalsolution}.\addra{ This is due to the fact that although $\widehat{\text{SNR}}_{\text{out}}$ is a biased estimate of $\text{SNR}_{\text{out}}$, at high output SNR the shrinkage factor $\left(1+1/\widehat{\text{SNR}}_{\text{out}}\right)^{-1}$ becomes arbitrarily close to the optimal shrinkage $\left(1+1/\text{SNR}_{\text{out}}\right)^{-1}$.}

\vspace{-0.4cm}
\subsection{Approximate MSE at Frequencies with Low SNR}\label{subsec:MSElowSNR}
\delra{\delra{Recall that $\text{SNR}_{\text{out}}=\left|\eta\right|^2$, hence i}\addra{I}n the low SNR regime\addra{, where $\text{SNR}_{\text{out}}=\left|\eta\right|^2\ll1$}\delra{ $|\eta|\ll1$. Accordingly}, we have the following theorem,}\addra{The MSE of the SW estimator at low SNRs is given in the following theorem,} whose proof appears in Appendix \ref{AppALowSNR}.
\begin{theorem}\label{theorem3}
\emph{At frequencies with low SNR, namely with $|\eta|\ll1$, the MSE \eqref{MSE_of_SW} \delra{is given by}\addra{attains the following approximation}}
\begin{equation}\label{MSE_of_SW_low_SNR_theorem3}
\begin{aligned}
\hspace{-0.1cm}\text{MSE}_{\sw}&=\left|X\right|^2+\rho\cdot\sigma_{\text{eff}}^2+\mathcal{O}\left(|\eta|\right)\\
	&=\left|X\right|^2\cdot\left(1+\frac{\rho}{\text{SNR}_{\text{out}}}\right)+\mathcal{O}\left(\sqrt{\text{SNR}_{\text{out}}}\right)\triangleq\varepsilon^2_{\text{low}},
\end{aligned}
\end{equation}
\emph{where the scalar $\rho\approx0.0464$ is defined \delra{explicitly in \eqref{definitionofrho}.}\addraRI{as
\begin{equation}\label{definitionofrhonew}
\rho =\frac12 \int_8^\infty \left[\frac{t}2+\sqrt{t^2/4-2t}-2\right]\frac{1}{2} e^{-\frac{t}{2}} \mathrm{d}t.
\end{equation}
}}
\end{theorem}


\addraRI{Combining \eqref{MSE_of_SW_low_SNR_theorem3} with the definition of $\sigma_{\text{eff}}^2$, \eqref{effnoise}, shows that at frequencies where $H[k]$ is close to zero, there is significant noise amplification, which in our method is attenuated by the multiplicative factor $\rho\approx0.0464\ll 1$. As an example, let us compare our noise suppression to the classical (unbiased) LS estimator at frequencies containing only noise, namely $X=0$. At such frequencies, our MSE is 
\begin{equation}\label{MSE_low_SNR_in_terms_of_noise_system_ratio}
\text{MSE}_{\sw}=\rho\cdot\sigma_{\text{eff}}^2,
\end{equation}
where the MSE of the LS estimator is 
\begin{equation}\label{MSEcomparetoLS}
\text{MSE}_{\LS}=\sigma_{\text{eff}}^2=\rho^{-1}\cdot\text{MSE}_{\sw}.
\end{equation}
Since $-10\log_{10}\rho\approx13$[dB], this is a significant noise suppression. Additional numerical comparisons demonstrating the gain relative to other estimators are provided in Section \ref{sec:simulresults}.}

Nevertheless, this ``defense mechanism" is of course limited due to the fact that $X$ is unknown. \delra{Indeed, the SW estimator, as any other estimator in this setup, cannot accurately distinguish between the presence of a dominant signal\addra{ (i.e., large mean and small variance)} and a  highly ``energized" realization of noise\addra{ (i.e., small mean and large variance)}, which \delra{is}\addra{are both} manifested as a large estimated output SNR \eqref{ZasSNRout}. Consequently, in the lat\addra{t}er case, $|Z|$ will be above the threshold, causing an erroneous estimate.}\addra{ A large observed value $|Z|$ may be due either to the presence of a strong signal, or to a large noise deviation added to a weak signal. The latter case with $|Z|$ above the threshold, yields an erroneous estimate.}

\delra{One can now also quantify}\addra{It is also informative to evaluate} the {\myfontb\emph{optimality gap}} from the MSE \eqref{MMSEexpression} of the unrealizable MMSE solution \eqref{optimalsolution} in the low SNR regime. \delra{which is }It is given approximately by
\begin{equation}\label{optimalitygapLOWSNR}
	\text{MSE}_{\sw}-\text{MMSE}\approx|X|^2\cdot\left(\frac{\text{SNR}_{\text{out}}+\rho(1+\text{SNR}^{-1}_{\text{out}})}{1+\text{SNR}_{\text{out}}}\right),
\end{equation}
neglecting the $\mathcal{O}\left(\sqrt{\text{SNR}_{\text{out}}}\right)$ term. It is readily verified that for any fixed noise level $\sigma_{\text{eff}}^2$, when $\text{SNR}_{\text{out}}\to0$, the optimality gap approaches $\rho\cdot\sigma_{\text{eff}}^2$. We thus see that the optimality gap is governed by the noise level, and relative to the efficient LS estimator, it is reduced by a factor $\rho$, in compliance with \eqref{MSEcomparetoLS}.
\vspace{-0.35cm}
\subsection{Approximate MSE at Frequencies with High SNR}\label{subsec:MSEhighSNR}
In the high SNR regime $|\eta|\gg1$, we have the following theorem, whose proof appears in Appendix \ref{AppAHighSNR}.
\begin{theorem}\label{theorem4}
\emph{At frequencies with high SNR, namely with $|\eta|\gg1$, the MSE \eqref{MSE_of_SW} is given by}
\begin{equation}\label{MSE_of_SW_high_SNR_theorem4}
\begin{aligned}
\text{MSE}_{\sw}&=(1-p)\cdot\left|X\right|^2+p\cdot\frac{|X|^2}{|\eta|^2}+\mathcal{O}\left(\frac{|X|^2}{|\eta|^4}\right)\\
&=(1-p)\cdot\left|X\right|^2+p\cdot\sigma_{\text{eff}}^2+\mathcal{O}\left(\frac{\sigma_{\text{eff}}^2}{\text{SNR}_{\text{out}}}\right)\triangleq\varepsilon^2_{\text{high}},
\end{aligned}
\end{equation}
where $p=Q_{1}\left(\sqrt{2\cdot\text{SNR}_{\text{out}}},2\sqrt{2}\right)$ as in \eqref{finalexpprob}.
\end{theorem}

\begin{figure*}[t]
	\centering
	\begin{subfigure}[b]{0.39\textwidth}
		\includegraphics[width=\textwidth]{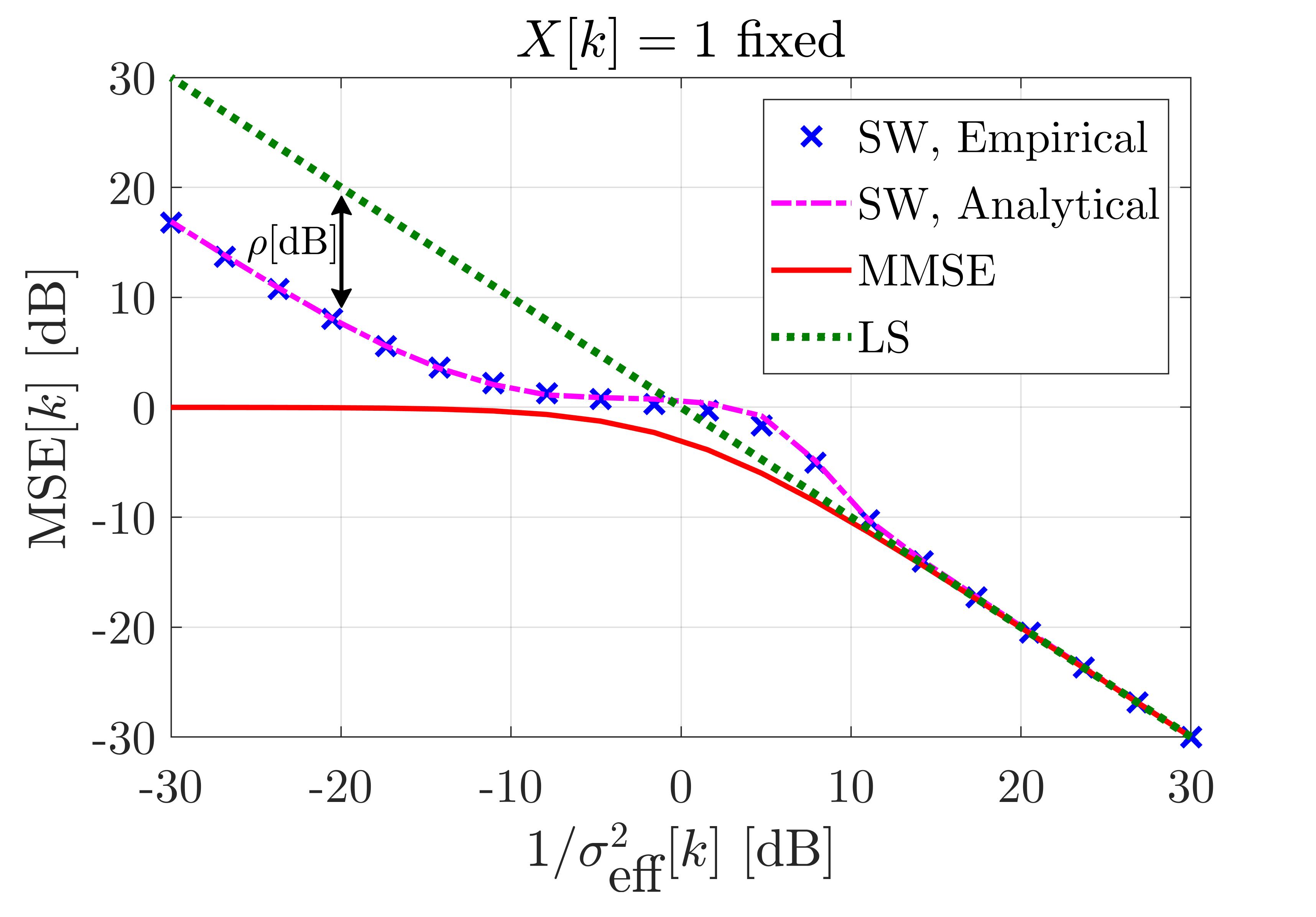}
		\caption{}\vspace{-0.1cm}
		\label{fig:predicted_MSE_X_fixed}
	\end{subfigure}%
	~ ~ ~ \qquad
	\begin{subfigure}[b]{0.39\textwidth}
		\includegraphics[width=\textwidth]{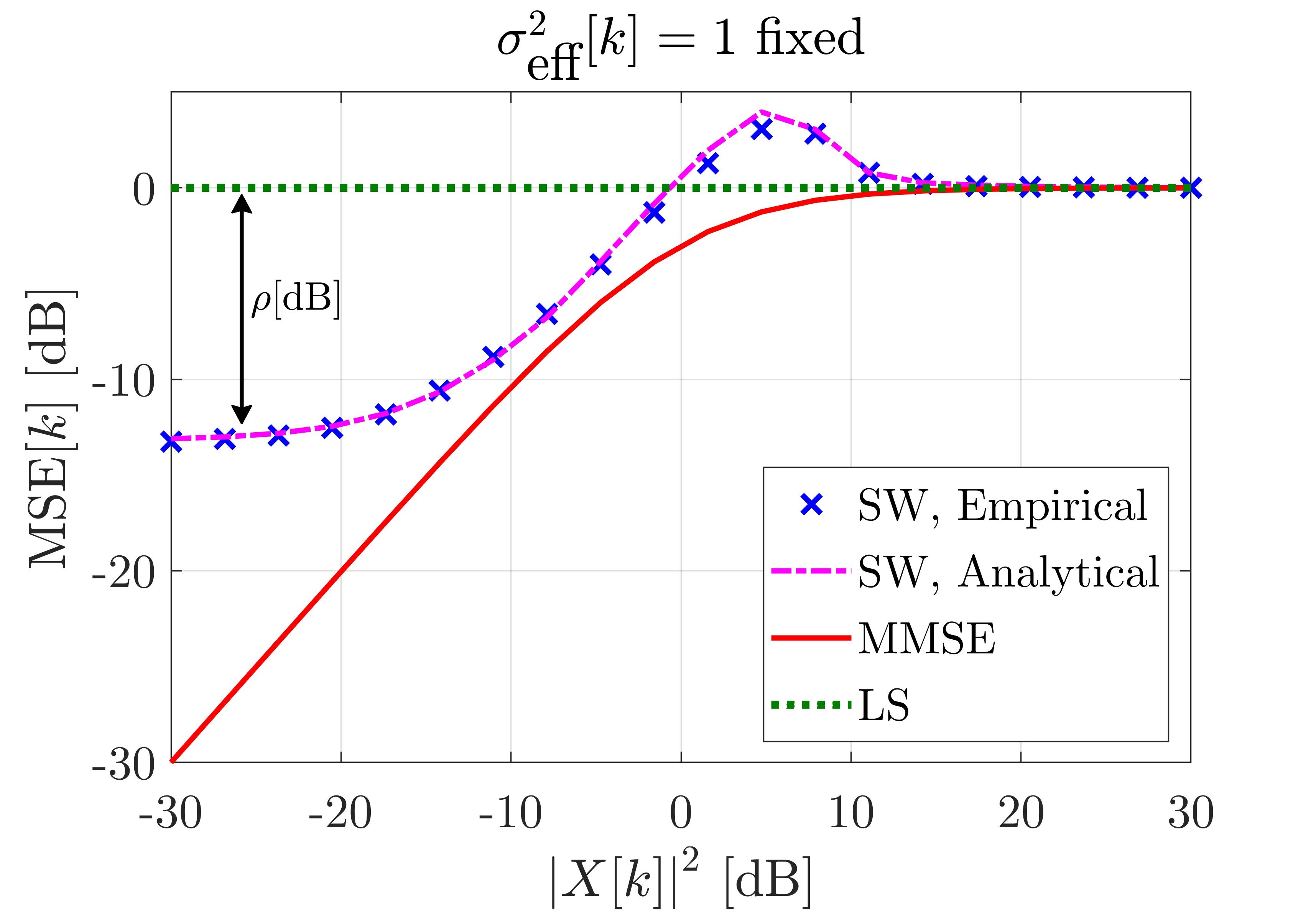}
		\caption{}\vspace{-0.1cm}
		\label{fig:predicted_MSE_sigma_v_fixed}
	\end{subfigure}
	\caption{The MSE of the $k$-th frequency component vs.\ (a) noise power (b) signal power. Here, $V[k]$ is zero-mean CN, and the empirical results were obtained by averaging $10^6$ independent trials. A good fit is seen between the analytical prediction and the empirical results. It is seen that in the low SNR regime the SW estimator reduces the MSE in $\sim\hspace{-0.05cm}13$ [dB] relative to the LS estimator, while retaining asymptotic optimality in the high SNR regime.}
	\label{fig:predicted_MSE}\vspace{-0.5cm}
\end{figure*}

\delra{In principle, from}\addra{Using} properties of\addra{ the} Marcum Q-function, as $|\eta|\to\infty$, $p\to1$ exponentially fast in $|\eta|$. When \delra{the output SNR}\addra{$\text{SNR}_{\text{out}}$} is sufficiently high and $p\approx1$ as in \eqref{marcumqhighsNR}, the optimality gap is\delra{ approximately} given by
\begin{equation}\label{optimalitygapHIGHSNR}
\hspace{-0.1cm}\text{MSE}_{\sw}\hspace{-0.05cm}-\hspace{-0.05cm}\text{MMSE}\approx\sigma_{\text{eff}}^2\delra{\cdot\frac{1}{1+\text{SNR}_{\text{out}}}}\addra{/\left(1+\text{SNR}_{\text{out}}\right)}.
\end{equation}
Following the discussion in Subsection \ref{subsec:highSNR}, \eqref{optimalitygapHIGHSNR} implies that the optimality gap approaches zero as \delra{the output SNR approaches infinity}\addra{$\text{SNR}_{\text{out}}\to\infty$},
\begin{equation}\label{optimalitygapHIGHSNRapprox}
\text{MSE}_{\sw}-\text{MMSE}\underset{\text{SNR}_{\text{out}}\gg1}{\approx}\delra{\frac{\sigma_{\text{eff}}^2}{\text{SNR}_{\text{out}}}}\addra{\sigma_{\text{eff}}^2/\text{SNR}_{\text{out}}}\xrightarrow[]{\text{SNR}_{\text{out}}{}\rightarrow\infty}0.
\end{equation}
We thus conclude that our proposed SW estimator converges to the optimal unrealizable solution as $\text{SNR}_{\text{out}}\rightarrow\infty$.

Having obtained approximated closed-form expressions for the MSE at low and high SNRs, for the intermediate SNR interval, we propose to interpolate between \eqref{MSE_of_SW_low_SNR_theorem3} and \eqref{MSE_of_SW_high_SNR_theorem4}. Specifically, for some fixed value $\tau\in\Rset^+$, we define
\begin{equation*}
f_{\tau}\left(\text{SNR}_{\text{out}}\right)\triangleq\frac{1}{2}\cdot\left(1-\frac{\text{SNR}_{\text{out}} \text{[dB]}}{\tau \text{[dB]}}\right), \forall \text{SNR}_{\text{out}}\in[-\tau,\tau] \text{[dB]},
\end{equation*}
with which the analytical approximation of the MSE {\myfontb\emph{per frequency}} of the SW estimator is given by
\begin{equation}\label{predictedMSEofSW}
\text{MSE}_{\sw}=\begin{cases}
\varepsilon^2_{\text{low}}, & \text{SNR}_{\text{out}}<-\tau \text{[dB]}\\
\varepsilon^2_{\text{low}}\cdot f_{\tau}+\varepsilon^2_{\text{high}}\cdot\left(1-f_{\tau}\right), & \hspace{-0.075cm}\left|\text{SNR}_{\text{out}}\right|\le\tau \text{[dB]}\\
\varepsilon^2_{\text{high}}, & \text{SNR}_{\text{out}}>\tau \text{[dB]}\\
\end{cases}.
\end{equation}
We emphasize that while the argument $\text{SNR}_{\text{out}}$ was omitted for brevity from the functions $\varepsilon^2_{\text{low}}, \varepsilon^2_{\text{high}}$ and $f_{\tau}$, they are all functions of $\text{SNR}_{\text{out}}$. Moreover, it is easily verified that the MSE \eqref{predictedMSEofSW} is a continuous function of $\text{SNR}_{\text{out}}$. In particular, $f_{\tau}(\text{SNR}_{\text{out}}=-\tau)=1$ and $f_{\tau}(\text{SNR}_{\text{out}}=\tau)=0$. Finally, based on our experience, a reasonable choice for $\tau$ is $6$ [dB].\addra{ For the derivation of the approximate MSE \eqref{predictedMSEofSW} in the intermediate SNR range\addraRI{ and further justification for the proposed interpolation, as well as the choice of $\tau$}, see Appendix \ref{AppAintermediateSNR}.}
\vspace{-0.2cm}
\section{Simulation Results}\label{sec:simulresults}
\vspace{-0.1cm}
In this section we first present empirical results that corroborate our analytical derivations regarding the predicted performance of the proposed SW estimator \eqref{sw_est}. Then, we compare our proposed method with four other methods for three different input signals. Within this simulated experiment, we demonstrate that our proposed SW estimator achieves good performance even when the constant noise spectral level is unknown and is estimated from the observed noisy convol\addra{v}\delra{ut}ed signal. \delra{Finally, w}\addra{W}e\addra{ then} also demonstrate the accuracy of our performance analysis for two cases of non-Gaussian noise, considering both heavy-tailed Laplace and compactly supported uniform distributed time-domain noise.\addra{ Finally, we consider the reconstruction of an ideal LIDAR signal corrupted by a smooth blur (e.g., \cite{johnstone2004wavelet}, Fig.\ 1 and Fig.\ 2a, and \cite{cavalier2007wavelet}, Fig.\ 1a), a widely used setting, considered as a benchmark for comparison of different deconvolution methods.}
\vspace{-0.3cm}
\subsection{Predicted MSE Per Frequency of the SW Estimator}
We \delra{now }examine the MSE {\myfontb\emph{per frequency}} of the proposed estimator. As can be seen from \eqref{MSE_of_SW_low_SNR_theorem3} and \eqref{MSE_of_SW_high_SNR_theorem4}, the resulting MSE does not depend on the phase of $X[k]$, nor of $H[k]$. In addition, the MSE \eqref{predictedMSEofSW} is a function of only two quantities: $|X[k]|^2$ and $\sigma_{\text{eff}}^2[k]=S_v[k]/|H[k]|^2$. Thus, in general, a surface plot is sufficient to fully describe the dependence of the MSE on $|X[k]|$ and $\sigma_{\text{eff}}^2[k]$. However, for enhanced visibility we present only two representative slices of this surface. Specifically, Fig.\ \ref{fig:predicted_MSE} \delra{presents the empirical and analytically predicted MSE \eqref{predictedMSEofSW} of the SW estimator, as well as the predicted MSEs of the LS estimator \eqref{LSestimate} and}\addra{ compares the empirical MSE of the SW estimator, its analytical approximation \eqref{predictedMSEofSW}, the MSE of the LS estimator, and for reference, also} the MSE of the (unrealizable) optimal solution\delra{ for reference}. \delra{While i}\addra{I}n Fig.\ \ref{fig:predicted_MSE_X_fixed} $X[k]=1$ is fixed and $\sigma_{\text{eff}}^2[k]$ is varied, \addra{whereas }in Fig.\ \ref{fig:predicted_MSE_sigma_v_fixed} $\sigma_{\text{eff}}^2[k]=1$ and we vary $X[k]$. The noise $V[k]$ was drawn from the circular CN distribution, and each point in the graph is the average of $10^6$ independent trials.

First, it is evident that the analytical\delra{ly predicted}\addra{ formula \eqref{predictedMSEofSW} for the} MSE\delra{ \eqref{predictedMSEofSW}} is in excellent fit with the empirical results, verifying that our \delra{analytical }analysis \delra{of the expected performance in terms of MSE }is fairly accurate. Second,\delra{as explained in Subsection \ref{subsec:MSElowSNR}, it can be seen that the MSE of our estimator depends on the unknown signal. In particular,} the noise-suppression mechanism is evident in the low SNR regime. This is in stark contrast to the LS estimator, whose MSE \eqref{MSEcomparetoLS} is independent of the signal, and is therefore $\rho$ [dB] higher than the MSE of the SW estimator at low SNRs. Finally, in compliance with \eqref{optimalitygapHIGHSNRapprox}, the performance of our estimator is asymptotically optimal as $\text{SNR}_{\text{out}}\rightarrow\infty$, similarly to the LS estimator.

Note that we do not assume to have prior knowledge on the unknown signal's DFT structure, thus each frequency component may have any arbitrary SNR. Therefore, a desirable estimator will provide good performance in terms of MSE (preferably) for any output SNR {\myfontb\emph{per frequency}}. Figs.\ \ref{fig:predicted_MSE_X_fixed} and \ref{fig:predicted_MSE_sigma_v_fixed} show that our proposed estimate has this property, where the ``price" paid for this overall SNR behavior is a local performance degradation in the intermediate SNR region, in which threshold-type estimators generally suffer the most.
\vspace{-0.3cm}
\subsection{Comparison to Other Deconvolution Methods}\label{subsec:simulationcomparisontoothermethods}

\begin{figure}[t]
	\centering
	\includegraphics[width=0.375\textwidth]{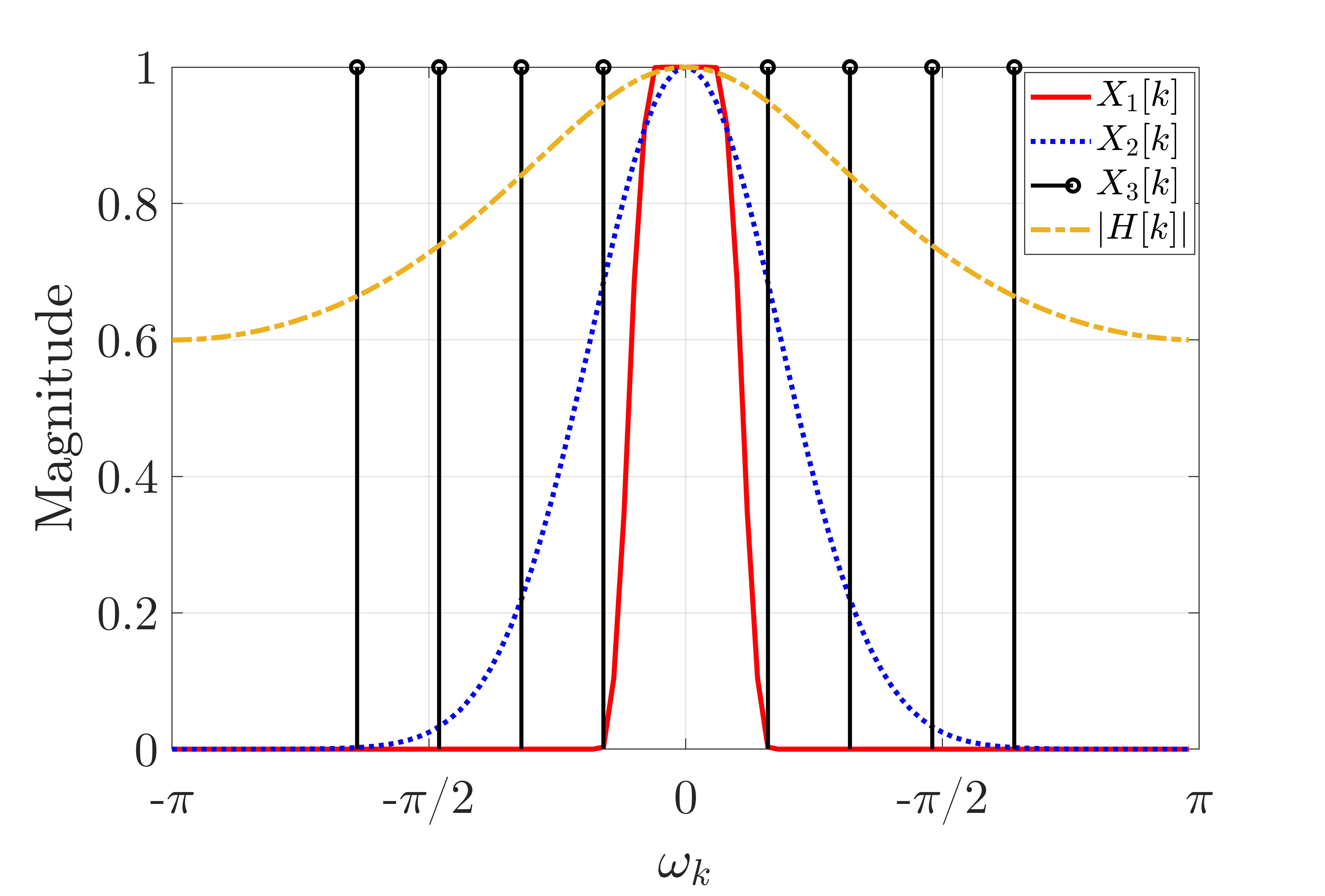}
	\caption{The DFT magnitudes of the unknown signals $\{X_i[k]\}_{i=1}^3$ and the known filter $|H[k]|$ with $\alpha=0.25$ for $N=100$. $X_1[k]$, bandlimited signal, $X_2[k]$, Gaussian pulse, $X_3[k]$, narrowband, frequency-domain sparse signal.}\vspace{-0.2cm}
	\label{fig:XandH}\vspace{-0.4cm}
\end{figure}
\begin{figure*}[t]
	\centering
	\begin{subfigure}[b]{0.32\textwidth}
		\includegraphics[width=\textwidth]{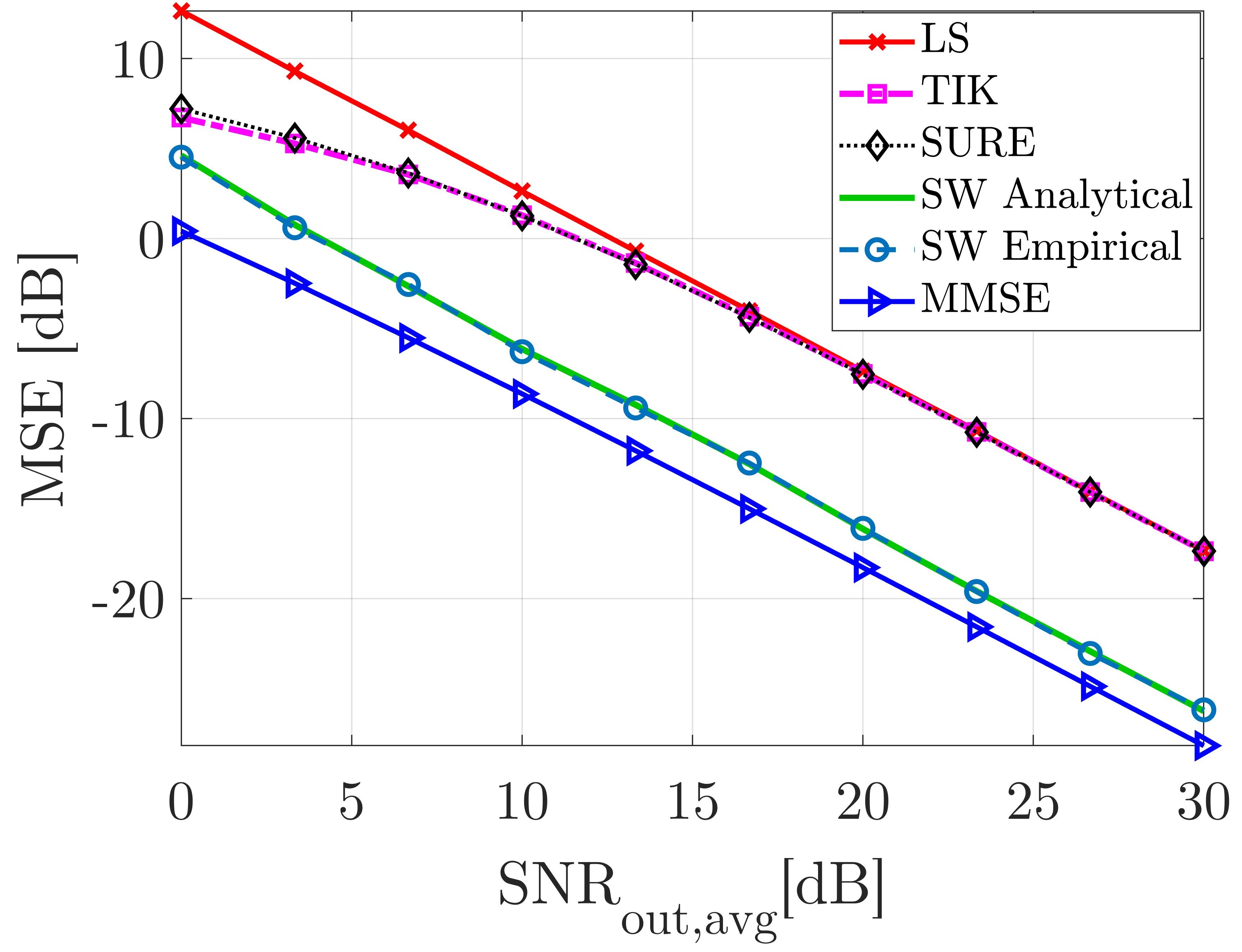}\vspace{-0.1cm}
		\caption{}\vspace{-0.1cm}
		\label{fig:MSE_vs_SNR_X1}
	\end{subfigure}%
	~
	\begin{subfigure}[b]{0.32\textwidth}
		\includegraphics[width=\textwidth]{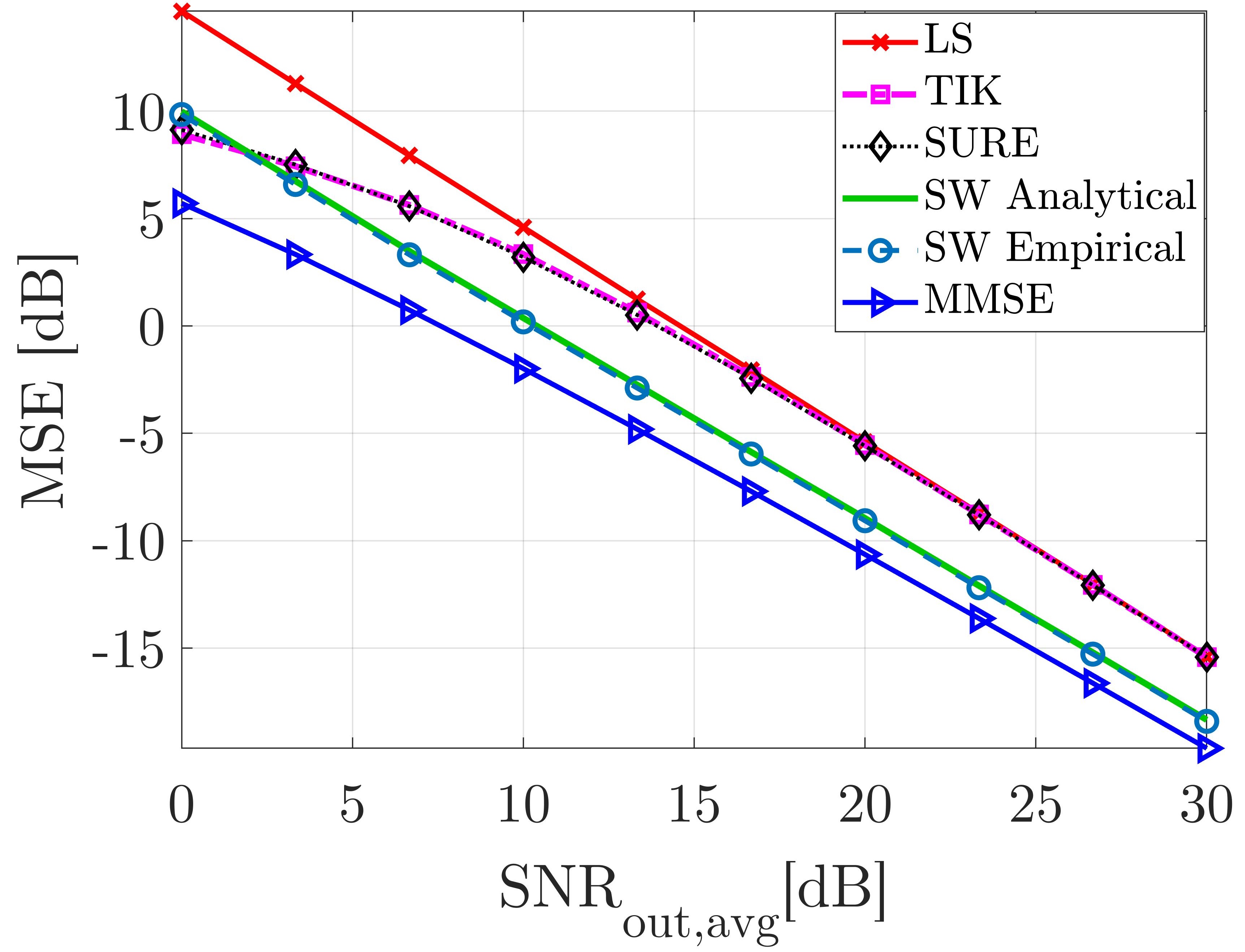}\vspace{-0.1cm}
		\caption{}\vspace{-0.1cm}
		\label{fig:MSE_vs_SNR_X2}
	\end{subfigure}
	~
	\begin{subfigure}[b]{0.32\textwidth}
		\includegraphics[width=\textwidth]{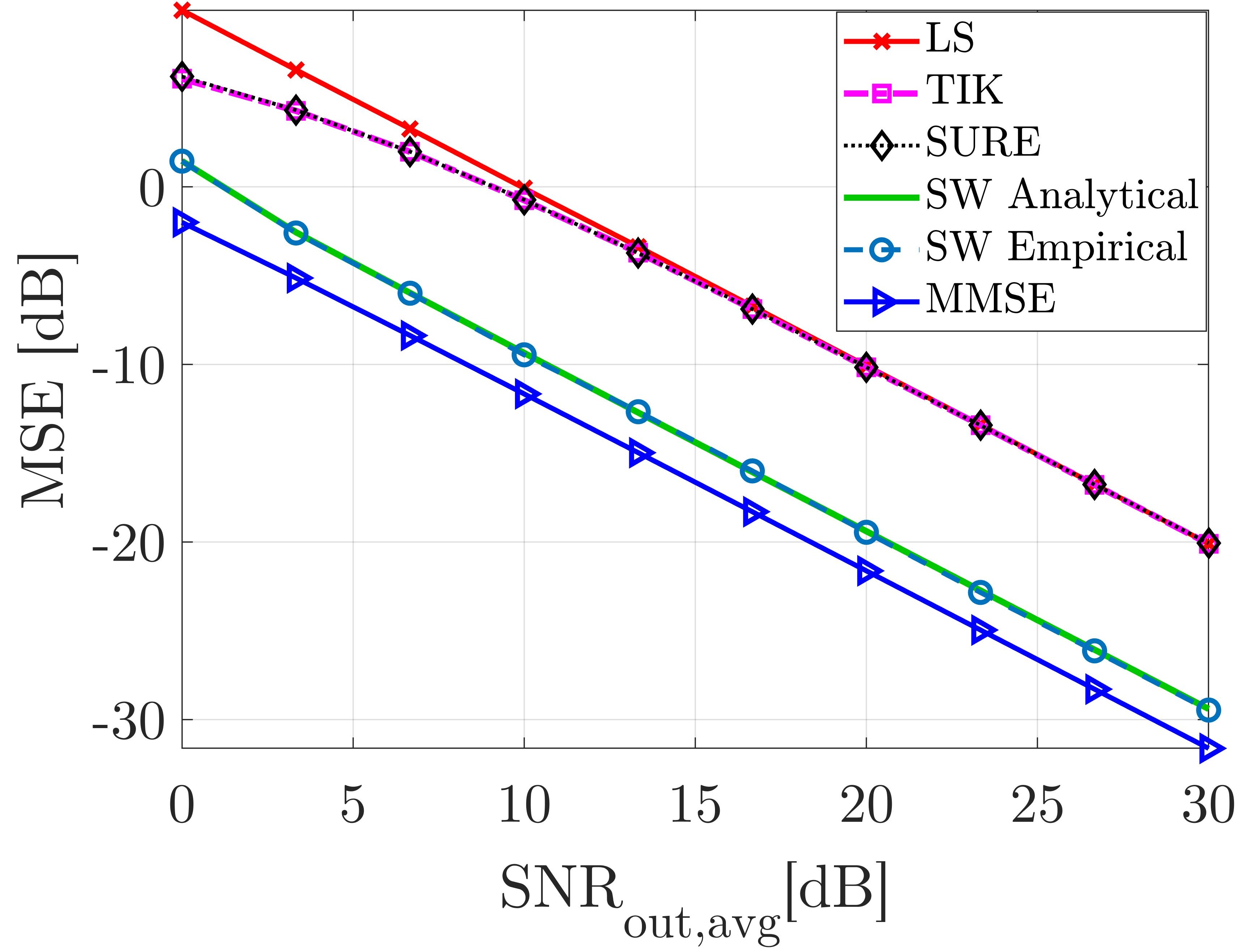}\vspace{-0.1cm}
		\caption{}\vspace{-0.1cm}
		\label{fig:MSE_vs_SNR_X3}
	\end{subfigure}
	\caption{MSE vs.\ the {\myfontb\emph{average}} output SNR. As seen, our estimator exhibits the best overall performance, and is the closest to the unrealizable optimal solution. Results were obtained by $10^3$ independent trials for $N=100$. (a) $X_1[k]$, bandlimited signal (b) $X_2[k]$, Gaussian pulse (c) $X_3[k]$, Narrowband signal.}
	\label{fig:MSE_vs_SNRout}\vspace{-0.2cm}
\end{figure*}
\begin{figure*}[t]
	\centering
	\begin{subfigure}[b]{0.32\textwidth}
		\includegraphics[width=\textwidth]{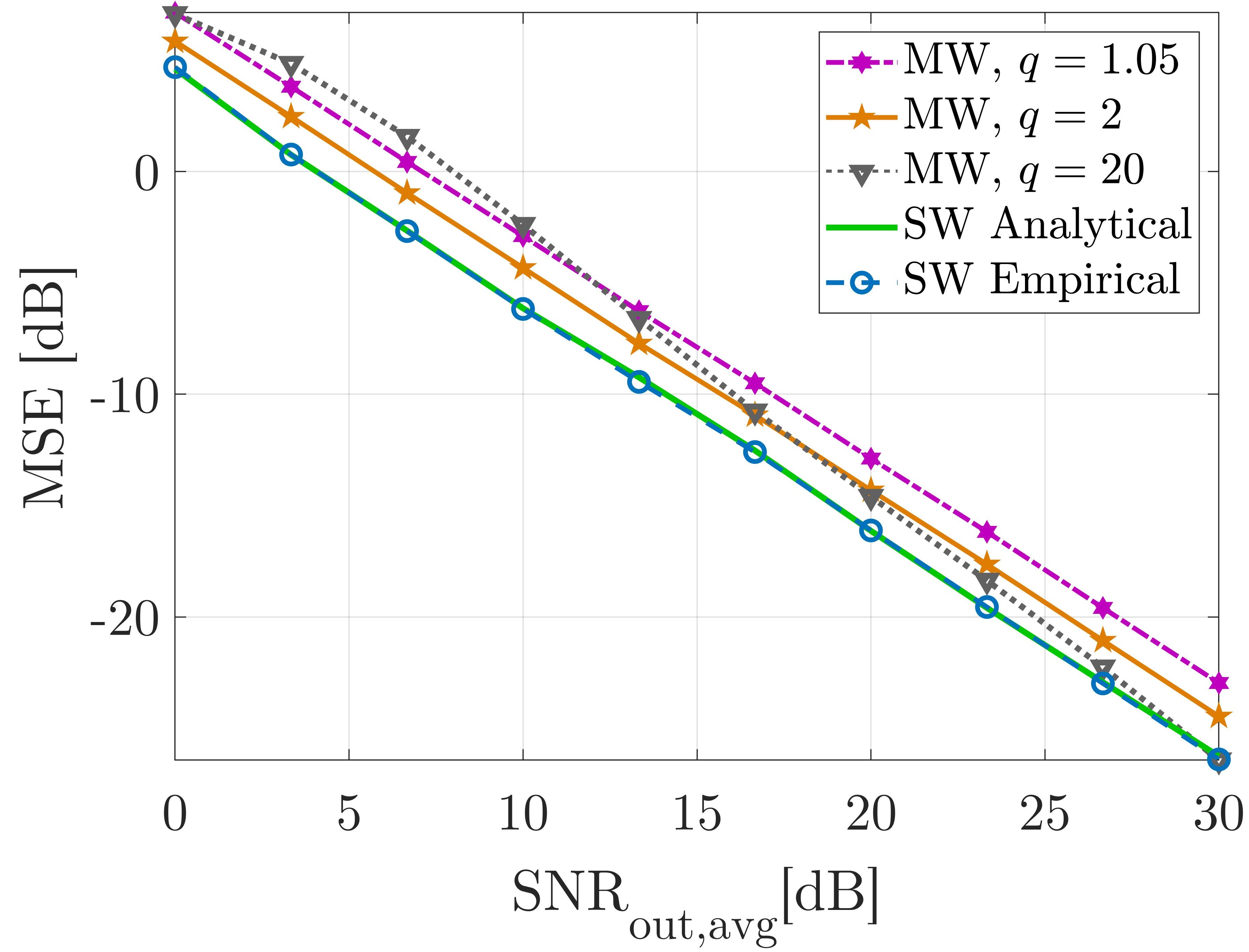}
		\caption{}
		\label{fig:MSE_vs_SNR_X1_MW}
	\end{subfigure}%
	~
	\begin{subfigure}[b]{0.32\textwidth}
		\includegraphics[width=\textwidth]{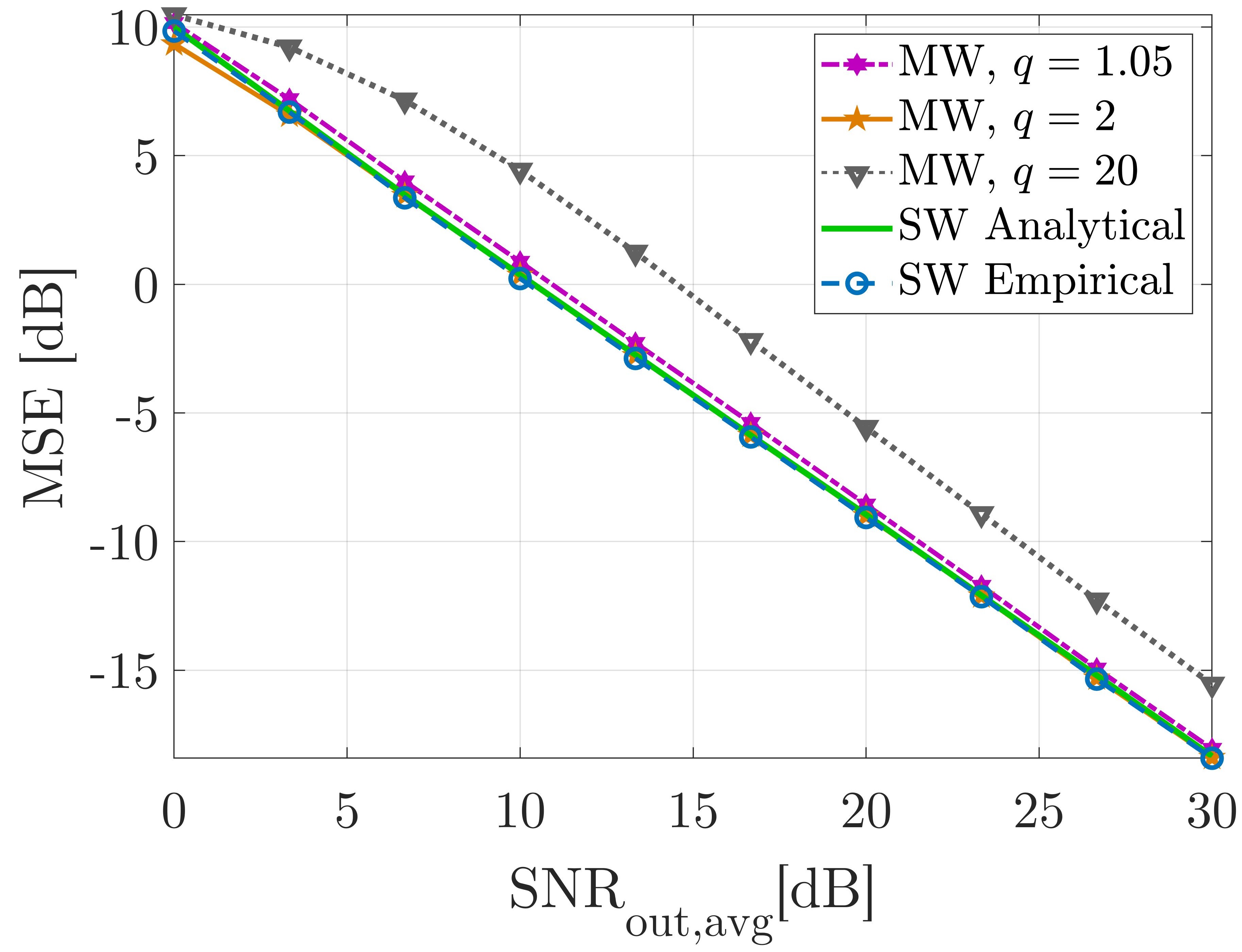}
		\caption{}
		\label{fig:MSE_vs_SNR_X2_MW}
	\end{subfigure}
	~
	\begin{subfigure}[b]{0.32\textwidth}
		\includegraphics[width=\textwidth]{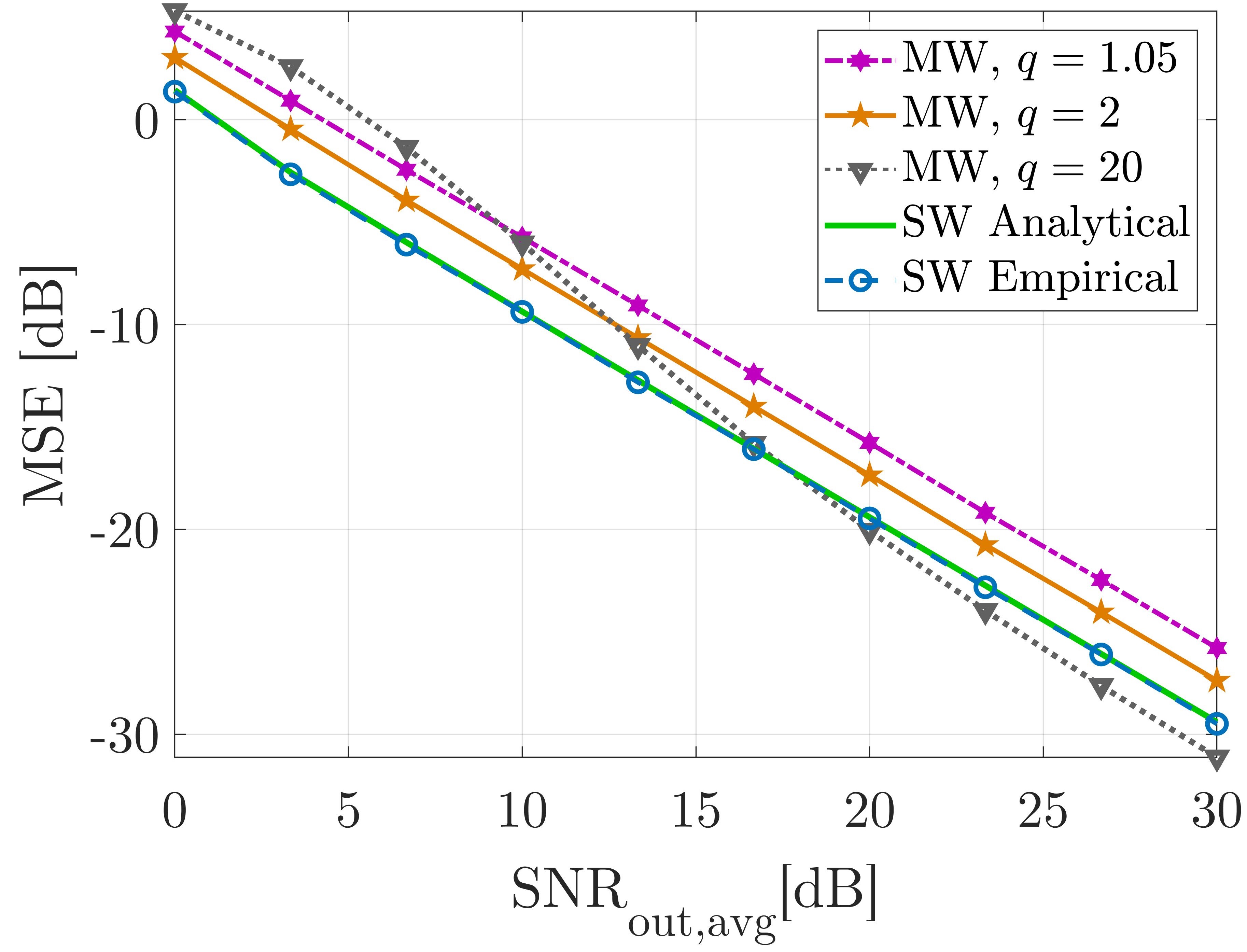}
		\caption{}
		\label{fig:MSE_vs_SNR_X3_MW}
	\end{subfigure}
	\caption{MSE vs.\ the {\myfontb\emph{average}} output SNR. While the MW estimator requires careful calibration of the tuning parameter $q$, and its performance are quite sensitive w.r.t.\ $q$, the proposed SW estimator requires no tuning and provides better performance overall. Results were obtained by $10^3$ independent trials for $N=100$. (a) $X_1[k]$, bandlimited signal (b) $X_2[k]$, Gaussian pulse (c) $X_3[k]$, Narrowband signal.}
	\label{fig:MSE_vs_SNRout_MW}
\end{figure*}
Consider the following three signals of length $N=100$,
\begin{align}
X_1[k]&=\left.\left[\tfrac{2\pi}{\Delta}\left(\text{rect}\left(\tfrac{\Omega}{2\pi}\right)*\text{tri}\left(\tfrac{\Omega}{\Delta}\right)\right)\right]\right|_{\Omega=6\Delta\omega_k},\label{signalsimulation1}\\
X_2[k]&=\left.\left[e^{-|\Omega|^2}\right]\right|_{\Omega=\sqrt{\Delta}\omega_k},\label{signalsimulation2}
\end{align}
\begin{align}
X_3[k]&=\sum_{\ell=1}^{4}{\left(\delta[k-8\ell]+\delta[k+8\ell]\right)},\label{signalsimulation3}
\end{align}
with $\Delta=1.5$,
and an LTI system with a frequency response
\begin{equation}\label{Frequencyresponse}
H[k]=\frac{1-\alpha}{1-\alpha\cdot e^{-\jmath\omega_k}}, \;\; \alpha\in(-1,1),
\end{equation}
where $\omega_k\triangleq\tfrac{2\pi}{N}k$, $*$ denotes continuous convolution, $\delta[\cdot]$ denotes Kronecker's delta, and $\text{rect}(\cdot), \text{tri}(\cdot)$ are the standard rectangular and triangular functions, resp\addraRI{ectively}. Notice that \eqref{signalsimulation1}, \eqref{signalsimulation2} and \eqref{signalsimulation3} are the DFTs of a bandlimited pulse, a Gaussian (approximately bandlimited) pulse, and a narrowband, frequency-domain sparse signal, resp\addraRI{ectively}. These functions are representative of common physical signals in various applications. Further, the frequency response \eqref{Frequencyresponse} corresponds to an FIR filter, approximating\footnote{In our case, for $N=100$, the approximation error is completely negligible: for $\alpha=0.25$, already at $n=50$ we have $h[50]\approx-2.3666\cdot10^{-30}$.} the infinite impulse response $h[n]=(1-\alpha)(-\alpha)^nu[n]$ via truncation, where $u[n]$ is the Heaviside step function. For negative values of $\alpha$, $H[k]$ is a non-ideal high pass filter, whereas for positive values, $H[k]$ is a non-ideal low pass filter. Here, we set $\alpha=0.25$, with which $H[k]$ has approximately the same effect as the smooth blur considered in \cite{johnstone2004wavelet} (see Subsection 2.1, Fig.\ 2a therein). For simplicity, we consider the case of white noise, thus the PSD of $v[n]$ is $S_v[k]=\sigma_v^2$ for all $k$. The magnitudes $\{|X_i[k]|\}_{i=1}^3$ and $|H[k]|$ with $\alpha=0.25$ are presented in Fig.\ \ref{fig:XandH}. Observe that with the signals \eqref{signalsimulation1}--\eqref{signalsimulation3} and the system \eqref{Frequencyresponse}, the values $\{|X_i[k]H[k]|^2\}$ range from $0$ to $1$ in (almost) all the range. Thus, by varying $\sigma_v^2$, the following empirical evaluation puts to test the considered deconvolution methods below in a very wide range of output SNR {\myfontb\emph{per frequency}}---from $-\infty$ [dB] to $\sim\hspace{-0.1cm}40$ [dB]---which fairly covers the output SNR range of practical interest.

We compare the MSE \eqref{costfunctionFreq} for the signals \eqref{signalsimulation1}--\eqref{signalsimulation3} and the system \eqref{Frequencyresponse}, achieved by the following five methods:
\begin{enumerate}[i.]
	\item The na\"ive LS  estimator (also known as inverse filter) \eqref{LSestimate};
	\item The grand-mean shrinkage of the Stein unbiased risk estimate type (SURE, \cite{xie2012sure}, Eq.\ (4.2));
	\item The Tikhonov (TIK) regularization-based \cite{tikhonov1977solutions},
	\begin{equation}\label{TIK_filter}
	G_{\text{TIK}}[k]\triangleq\frac{H^*[k]}{\left|H[k]\right|^2+\frac{\sigma_v^2}{P_x}}
	\end{equation}
	where $P_x\triangleq\tfrac{1}{N}\sum_{k=0}^{N-1}{\left|X[k]\right|^2}\in\Rset^+$ is the (known or estimated) average power of the unknown input signal;
	\item The minimum average MSE Modified Wiener (MW, \cite{walden1988robust}, Eq.\ (9)),
	\begin{equation}\label{MW_filter}
	G_{\text{MW}}[k]\triangleq\frac{1}{H[k]}\cdot\frac{1}{1+q\cdot\frac{1}{\widehat{\text{SNR}}_{\text{out,i}}[k]}},
	\end{equation}
	where $q$ is an adjustable noise-control parameter and $\widehat{\text{SNR}}_{\text{out,i}}[k]\triangleq\max\left(\widehat{\text{SNR}}_{\text{out}}[k]-1,0\right)$ is an improved output SNR estimator\footnote{Note that an output SNR estimate is not specified in \cite{walden1988robust}. Nevertheless, for MW, we use the improved, less biased estimate $\widehat{\text{SNR}}_{\text{out,i}}[k]$.} since $\Eset\left[\widehat{\text{SNR}}_{\text{out}}[k]\right]=\text{SNR}_{\text{out}}[k]+1$;
	\item Our proposed SW estimator \eqref{sw_est}.
\end{enumerate}
Note that LS and SURE are fully data-driven estimators. Further, note that TIK \eqref{TIK_filter} is the optimal filter \eqref{optimalsolution} for an input signal with a constant DFT, i.e., $\left|X[k]\right|^2=P_x$ for all $k\in\{0,\ldots,N-1\}$. However, for signals with a non-constant DFT, in practice, since $P_x$ is unknown, the regularization constant $(\sigma_v^2/P_x)$, sometimes termed the ``Tikhonov parameter", has to be tuned \textit{ad-hoc}. Specifically, in our simulations we consider an ideal (or ``oracle") TIK estimate, which enjoys the advantage of available side information---the exact power of the unknown input signal $x[n]$. As for MW \eqref{MW_filter}, since its performance is highly sensitive to the particular choice of the tuning parameter $q$, a more detailed comparison for a few values of $q$, along with a short discussion, will be shortly provided separately. Finally, as a benchmark for the lowest attainable MSE, we add \addra{the MSE of}\delra{to our simulations} $\widehat{X}_{\text{opt}}[k]$ of \eqref{optimalsolution}. This estimator is of course unrealizable, however it is still the optimal MMSE deconvolution-type \eqref{solutionform} solution. The following results were obtained by averaging $10^3$ independent trials.
\begin{figure*}[t]
	\centering
	\begin{subfigure}[b]{0.32\textwidth}
		\includegraphics[width=\textwidth]{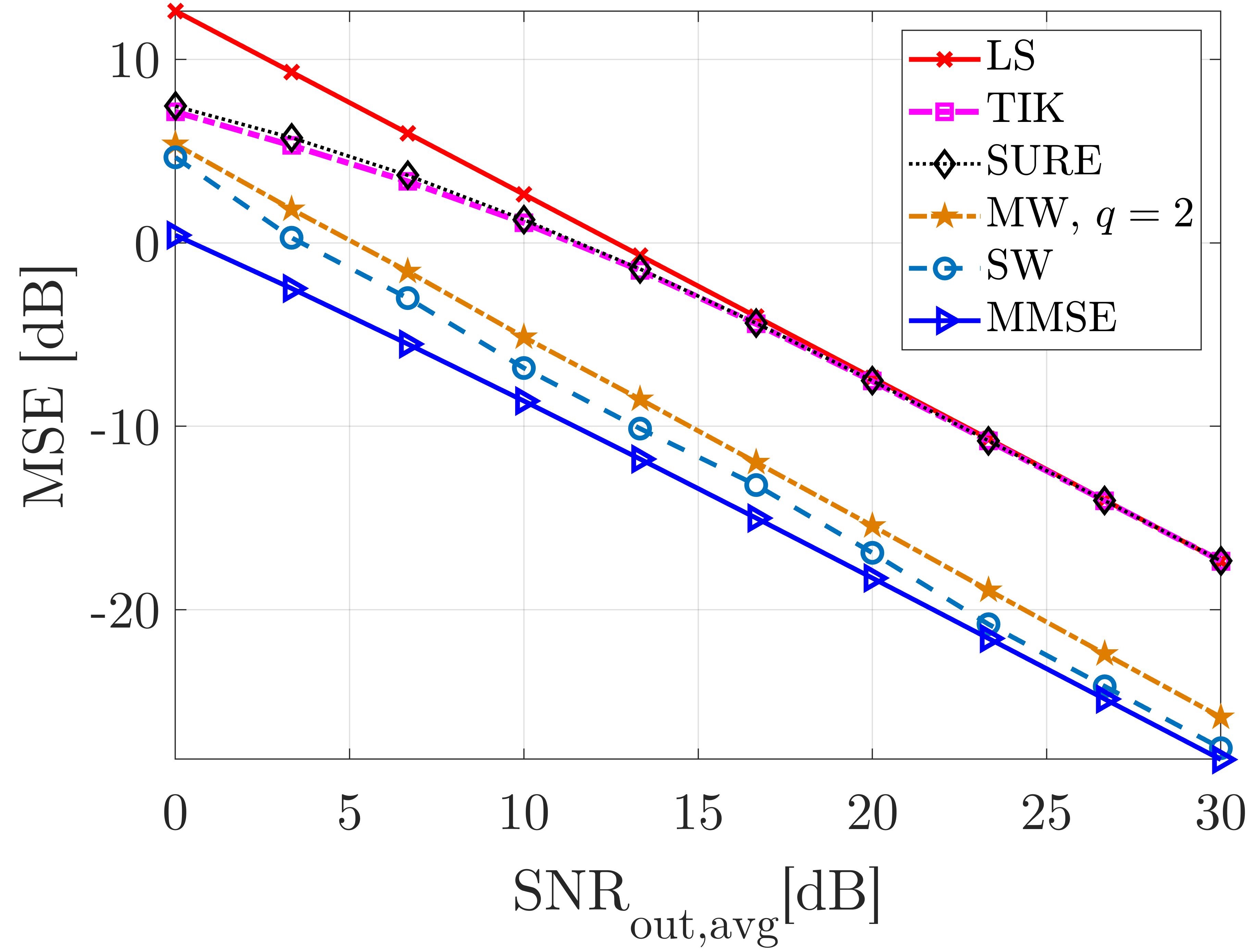}\vspace{-0.2cm}
		\caption{}\vspace{-0.2cm}
		\label{fig:MSE_vs_SNR_X1_estimated_noise}
	\end{subfigure}%
	~
	\begin{subfigure}[b]{0.32\textwidth}
		\includegraphics[width=\textwidth]{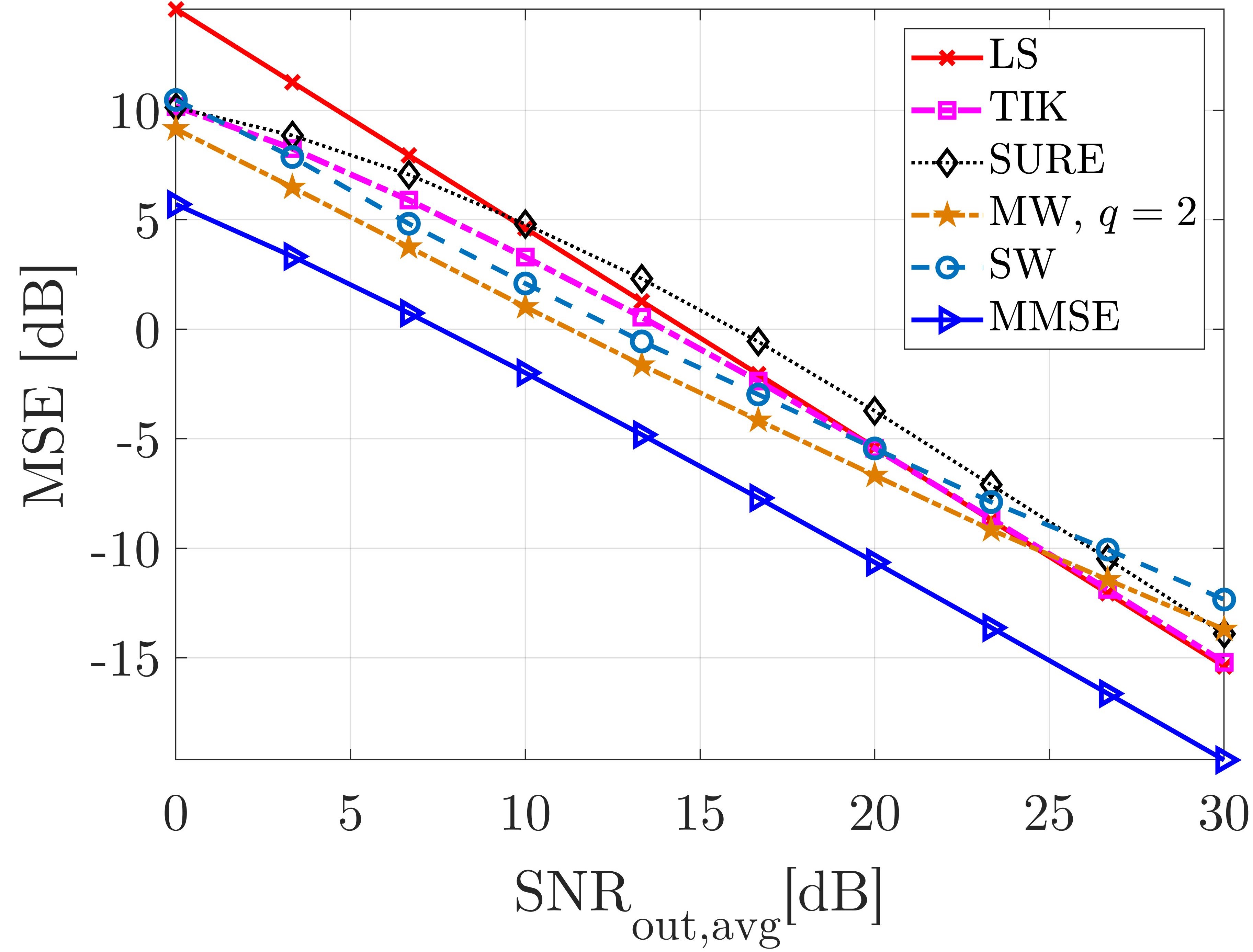}\vspace{-0.2cm}
		\caption{}\vspace{-0.2cm}
		\label{fig:MSE_vs_SNR_X2_estimated_noise}
	\end{subfigure}
	~
	\begin{subfigure}[b]{0.32\textwidth}
		\includegraphics[width=\textwidth]{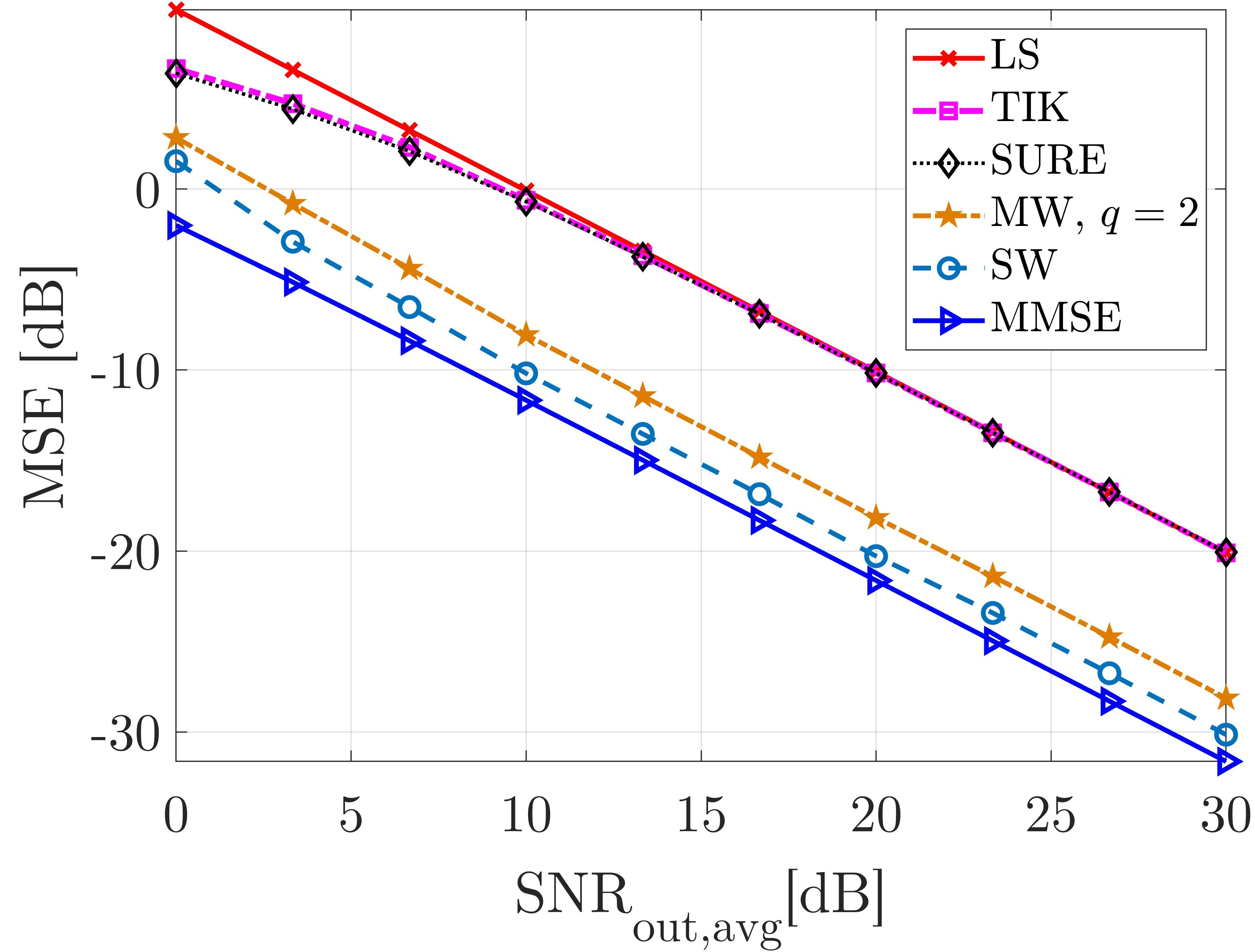}\vspace{-0.2cm}
		\caption{}\vspace{-0.2cm}
		\label{fig:MSE_vs_SNR_X3_estimated_noise}
	\end{subfigure}
	\caption{MSE vs.\ the {\myfontb\emph{average}} output SNR, where the noise variance $\sigma_v^2$ is unknown, and its estimate $\widehat{\sigma}_v^2$ is used instead. Results were obtained by averaging $10^3$ independent trials. (a) $X_1[k]$, bandlimited signal (b) $X_2[k]$, Gaussian pulse (c) $X_3[k]$, Narrowband signal. It is seen that our estimator exhibits the best performance for $X_1$ and $X_3$, whereas for $X_2$, different estimators dominate all the others in different SNR regions.}
	\label{fig:MSE_vs_SNRout_estimated_noise}\vspace{-0.4cm}
\end{figure*}
\begin{figure}[t]
	\centering
	\includegraphics[width=0.4\textwidth]{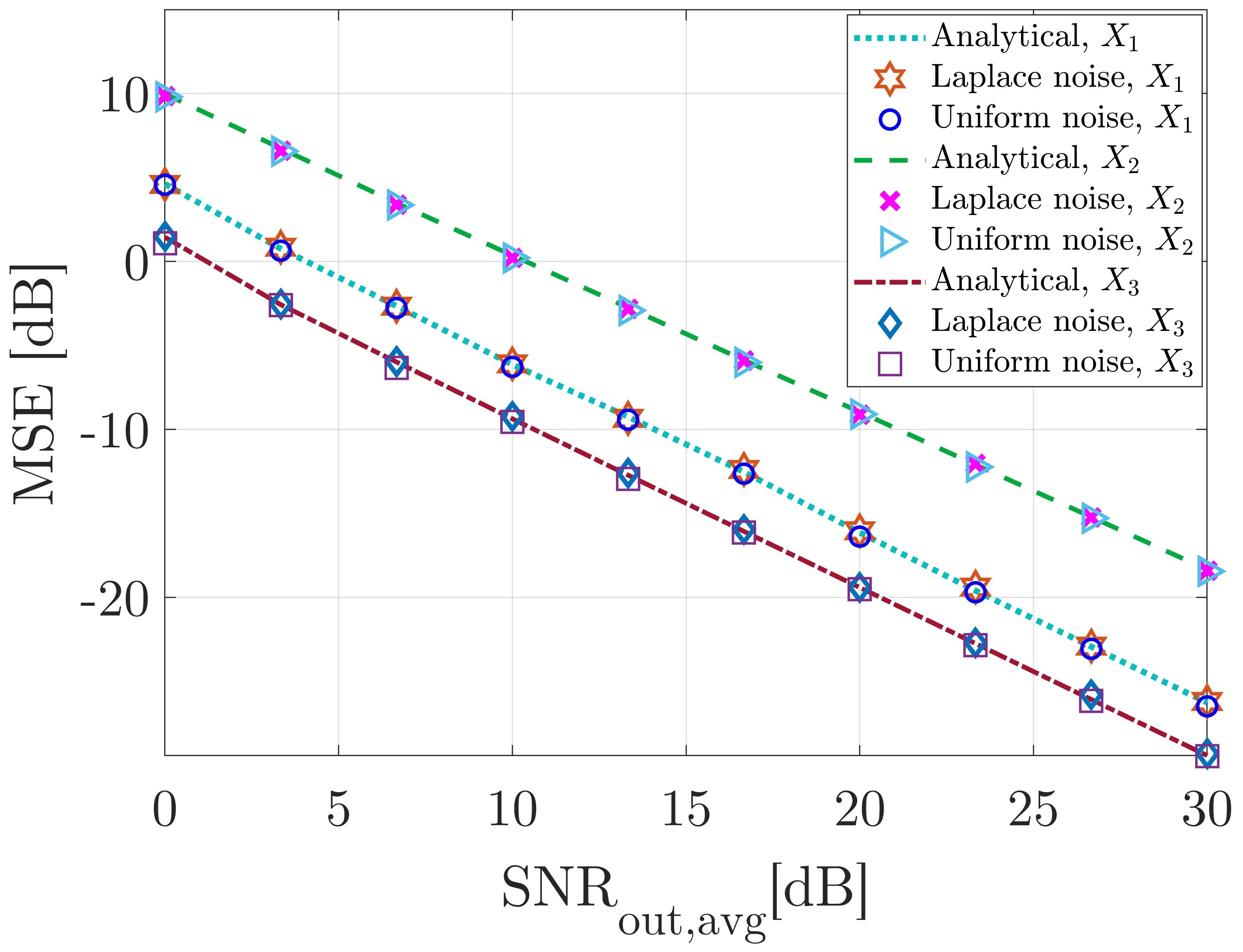}
	\caption{MSE of the SW estimator vs.\ the average output SNR for the signals $X_1, X_2$ and $X_3$ and non-Gaussian noise. Evidently, the CN distribution assumption for the noise {\myfontb\emph{\addraRI{per} frequency}} $V[k]$ enables to accurately assess the predicted MSE. Results were obtained by averaging $10^3$ independent trials.}
	\label{fig:nonGaussiannoise}\vspace{-0.4cm}
\end{figure}
\addra{\begin{figure}[t]
	\centering
	\begin{subfigure}[b]{0.225\textwidth}
		\includegraphics[width=\textwidth]{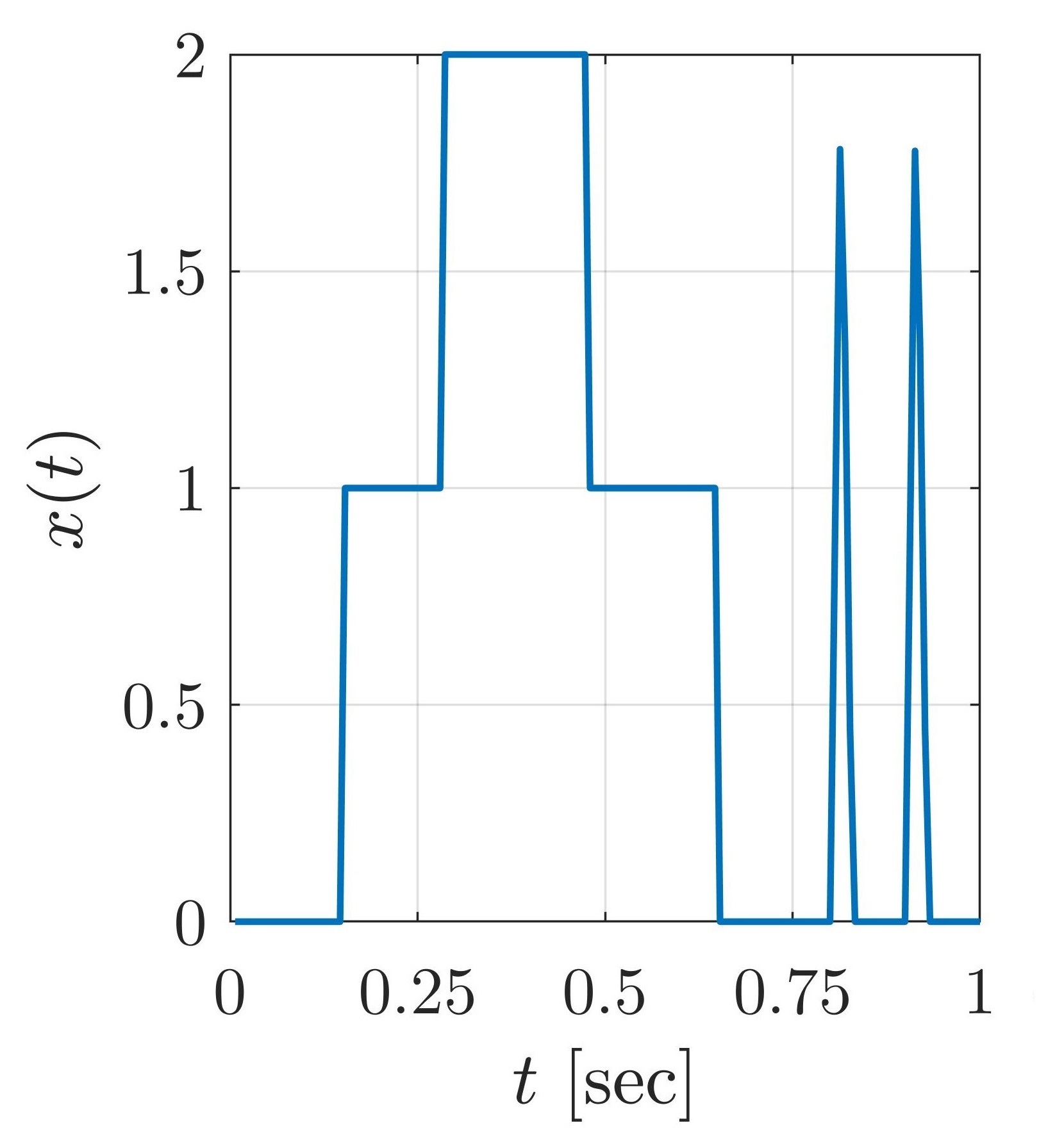}\vspace{-0.2cm}
		\caption{}\vspace{-0.1cm}
		\label{fig:LIDAR}
	\end{subfigure}
	\begin{subfigure}[b]{0.225\textwidth}
		\includegraphics[width=\textwidth]{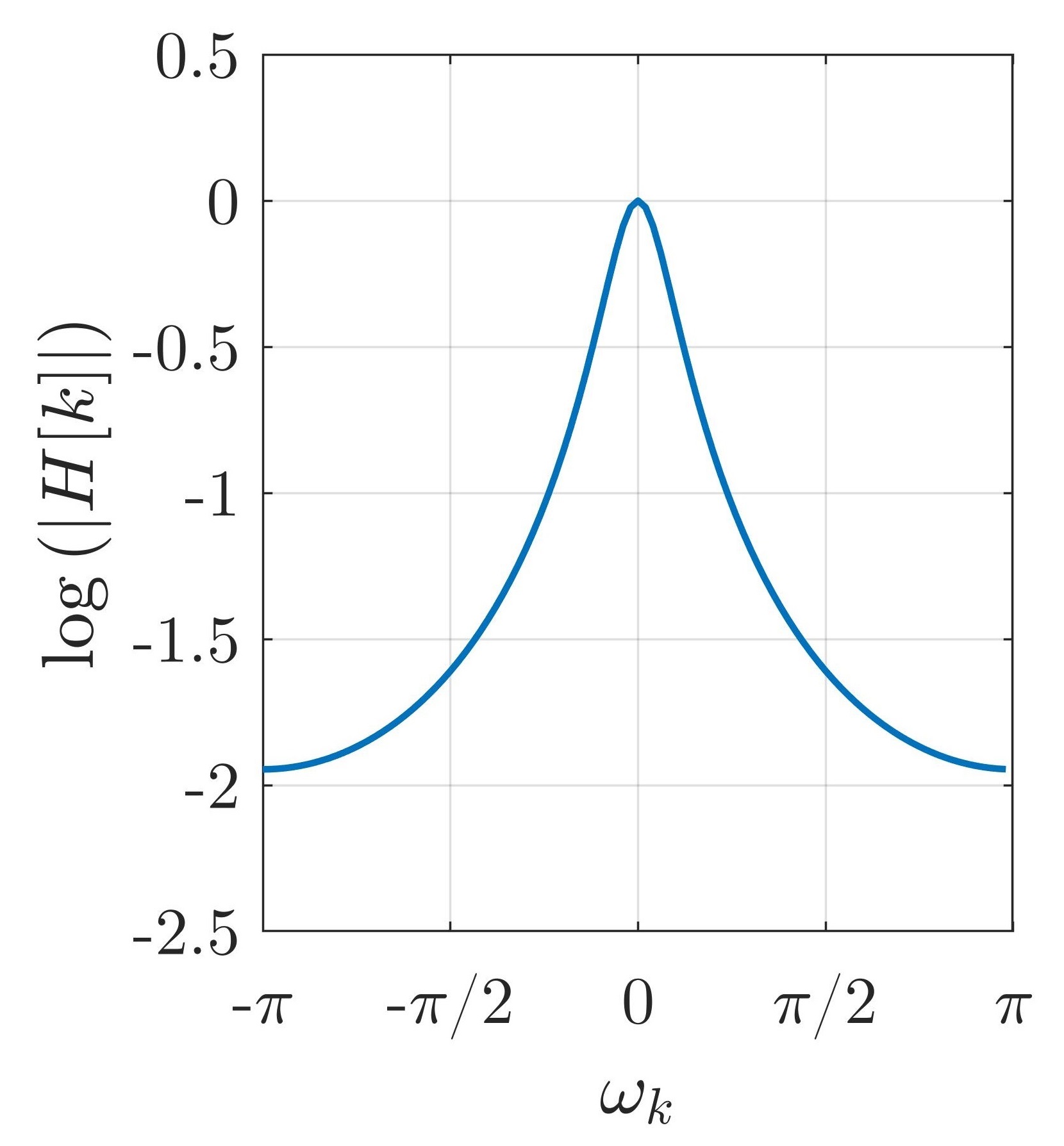}\vspace{-0.2cm}
		\caption{}\vspace{-0.1cm}
		\label{fig:smoothblur}
	\end{subfigure}
	\caption{(a) Time-domain ideal LIDAR signal (unknown input). (b) Log-magnitude of the known frequency response (``smooth blur").}
	\label{fig:LIDARandBLUR2}\vspace{-0.4cm}
\end{figure}
\begin{figure}[t]
	\centering
	\includegraphics[width=0.37\textwidth]{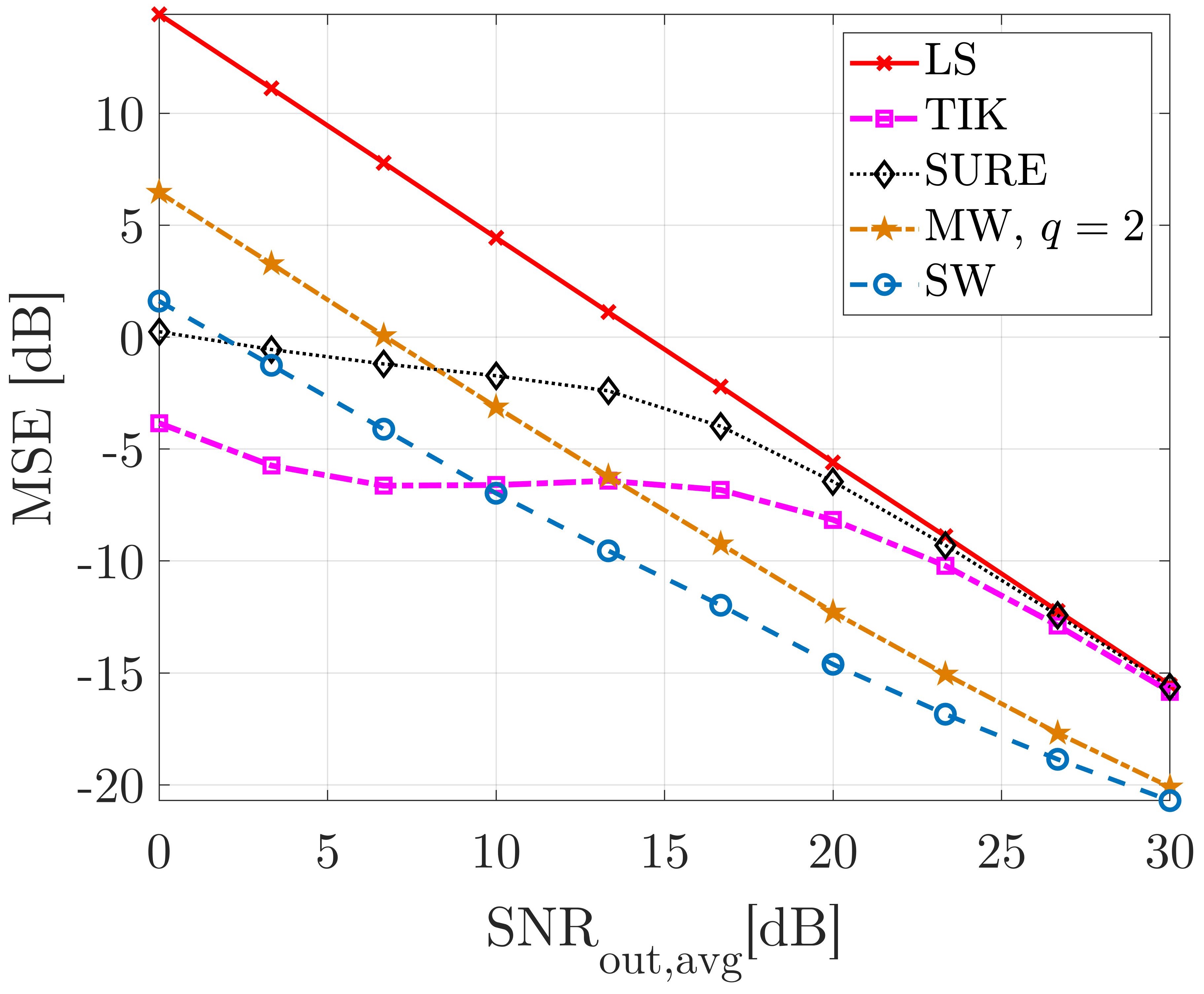}
	\caption{\addra{MSE, obtained by $10^3$ realizations, vs.\ the average output SNR in reconstruction of the LIDAR signal from its noisy, blurred measured version}}
	\label{fig:MSE_vs_SNR_out_LIDAR}\vspace{-0.5cm}
\end{figure}
}

Figure \ref{fig:MSE_vs_SNRout} presents the MSE \eqref{costfunctionFreq} vs.\ the {\myfontb\emph{average}} output SNR over all the output per-frequency SNRs, defined as
\begin{equation}\label{averageoutputSNR}
\text{SNR}_{\text{out,avg}}\triangleq\frac{1}{N}\sum_{k=0}^{N-1}{\text{SNR}_{\text{out}}[k]}.
\end{equation}
Evidently, our proposed estimator achieves a considerable improvement in the resulting MSE relative to the LS, TIK and SURE estimators, reaching a gain of almost an order of magnitude (up to $\sim\hspace{-0.05cm}9$ [dB]). Moreover, it is seen from Figs.\ \ref{fig:MSE_vs_SNR_X1} and \ref{fig:MSE_vs_SNR_X3} that our proposed estimator offers a more significant enhancement relative to these non-threshold-type methods for the bandlimited and sparse frequency-domain signals. In addition, although at the low SNR regime it is slightly inferior, our estimator is also superior to the other methods even for a non-sparse, only approximately bandlimited signal \eqref{signalsimulation2} for a sufficiently high {\myfontb\emph{average}} output SNR (in this example from $\sim\hspace{-0.05cm}2$ [dB]), as evident from Fig.\ \ref{fig:MSE_vs_SNR_X2}.

Next, we compare our proposed estimator with MW \eqref{MW_filter}, which depends on the tunning parameter $q$. Note that $q$ is fixed w.r.t.\ the different frequencies (i.e., the index $k$). Thus, we evaluate its performance for $q=1.05,2,20$, so as to examine different trade-offs between ``the mean-squared estimation error and the mean-squared filtered noise" (see \cite{walden1988robust}, Eq.\ (4)), referred here as MSE and noise suppression, resp\addraRI{ectively}\delra{.}:
\begin{itemize}
\item  With $q=1.05$, the MW tends to naively mimic the optimal solution \eqref{optimalsolution} ($\sim\hspace{-0.075cm}5\%$ noise suppression weight);
\item  With $q=2$, equal weights are given to MSE minimization and noise suppression; and
\item  With $q=20$, noise suppression is preferred over accurate signal reconstruction ($\sim\hspace{-0.075cm}95\%$ noise suppression weight).
\end{itemize}

Indeed, for many possible signals in various applications, while at some frequencies the SNR is very low or even zero, at others it may be very high, thus both noise suppression and MSE minimization are desired. As seen from Fig.\ \ref{fig:MSE_vs_SNRout_MW}, relative to the MW with $q=1.05$ and $q=2$, corresponding to an approximate na\"ive imitation of the optimal solution \eqref{optimalsolution} and equal weighting, resp\addraRI{ectively}\delra{.}, the SW estimator is uniformly superior for $X_1$ and $X_3$, and performs approximately equal\delra{it}\addra{l}y well for $X_2$. Further, the local superiority of the MW with $q=20$, corresponding to noise suppression oriented weighting, for $X_3$ at the high SNR regime {\myfontb{\emph{only}}}, is at the cost of greater degradation in the low SNR regime, and uniform inferiority to the SW estimator for $X_1$ and $X_2$. It is important to bear in mind that, in practice, since the input signal is unknown, and therefore the {\myfontb\emph{per frequency}} SNRs are unknown as well, the MSE cannot be evaluated, hence it is not clear how one chooses\footnote{\cite{walden1988robust} does not provide a method for choosing the tuning parameter $q$, but rather only discusses the effects of choosing different values of $q$.} the tuning parameter $q$, which clearly affects the resulting performance considerably. In contrast, since our proposed SW estimator is free of such a tuning parameter, in some sense, it implicitly chooses the proper ``weighting", according to the {\myfontb\emph{per frequency}} estimated output SNR \eqref{ZasSNRout}.

Unlike the previous comparison to LS, TIK and SURE in Fig.\ \ref{fig:MSE_vs_SNRout}, which emphasized the performance gain in terms of MSE, this comparison to MW emphasizes the inherent {\myfontb\emph{adaptivity}} property of our proposed solution. Accordingly, as also seen from Fig.\ \ref{fig:MSE_vs_SNRout_MW}, none of the three different MW estimators perform better than our proposed solution for all three signals. Instead, the SW estimator is the most stable, and exhibits the best overall performance, considering different signals with different average output SNRs.

Next, we compare the methods under a setting where the noise level $\sigma_v^2$ is unknown and has to be estimated. Following Donoho \etal\ \cite{donoho1995wavelet} (Subsection 6.1.1.), we use the Median Absolute Deviation (MAD) \cite{hampel1974influence,rousseeuw1993alternatives,leys2013detecting} to estimate $\sigma_v$,
\begin{equation}\label{noisevarianceestimate}
\widehat{\sigma}_v\triangleq\frac{1}{\sqrt{2}}\cdot1.4826\cdot\big[\text{MAD}\left(\Re\left\{\Y\right\}\right)+\text{MAD}\left(\Im\left\{\Y\right\}\right)\big].
\end{equation}
Here $\Y\triangleq\left[Y[0] \cdots Y[N-1]\right]^{\tps}\in\Cset^{N\times1}$, and as discussed in \cite{donoho1995wavelet}, \eqref{noisevarianceestimate} is accurate when the input signal is approximately sparse in the frequency domain. Accordingly, $\widehat{\sigma}_v^2$ replaces $\sigma_v^2$ for all methods. In particular, we have $\widehat{Z}[k]\triangleq Y[K]/\widehat{\sigma}_v\delra{^2}$ instead of $Z[k]$ for the SW estimator, and for a fair comparison, TIK now uses the estimated signal power
\begin{equation*}\label{signalpower4TIK}
\widehat{P}_x\triangleq\frac{1}{N}\sum_{k=0}^{N-1}|Y[k]|^2-\widehat{\sigma}_v^2\triangleq\widehat{P}_y-\widehat{\sigma}_v^2,
\end{equation*}
rather than the true $P_x$. For MW, we choose $q=2$, which is the most stable for the signals under consideration. The LS estimator simply applies the inverse filter and does not use $\sigma_v^2$. Thus, its performance is exactly the same as in the previous setting, where $\sigma_v^2$ is known (Fig.\ \ref{fig:MSE_vs_SNRout}).

Note that in our setting, where no assumptions on the input signal's DFT $X[k]$ are made, \eqref{noisevarianceestimate} is generally biased and overestimated. Therefore, the estimated output SNR \eqref{ZasSNRout} will now be lower. In turn, this will cause performance degradation at frequencies with high and intermediate SNR, since a higher shrinkage value \eqref{sw_shrinkage_form} will be wrongfully used. However, at frequencies with low SNR, the noise ``defense mechanism" discussed in Subsection \ref{subsec:MSElowSNR} will be intensified, and will result in performance enhancement. Thus, the overall deviation in the MSE \eqref{costfunctionFreq} depends on the true, {\myfontb\emph{unknown}} output SNR distribution over all frequencies. For example, bandlimited and/or sparse frequency-domain signals, whose majority of frequencies have low output SNR, are expected to have enhanced, or at least not degraded, overall MSE performance.

Figure \ref{fig:MSE_vs_SNRout_estimated_noise} shows the MSE \eqref{costfunctionFreq} vs.\ the {\myfontb\emph{average}} output SNR when $\widehat{\sigma}_v^2$ replaces $\sigma_v^2$. As seen, our estimator exhibits the best performance for $X_1$ and $X_3$, the bandlimited and sparse signals, resp\addraRI{ectively}. Further, a slight improvement up to $\sim\hspace{-0.075cm}1$ [dB] w.r.t.\ the previous setting is also observed, as expected, due to overestimation of $\sigma_v^2$. For $X_2$, which is not sparse or bandlimited, our proposed estimator exhibits performance degradation of up to $\sim\hspace{-0.075cm}6$ [dB] w.r.t.\ the previous setting in which $\sigma_v^2$ is known. Yet, it is still competitive, as different estimators dominate in different {\myfontb\emph{average}} output SNR regions. Therefore, our proposed method provides reliable deconvolution even when the unknown noise variance is estimated from the observed convolved signal itself.
\addraRI{
	\begin{figure*}[t]
		\centering
		\begin{subfigure}{\textwidth}
			\includegraphics[width=\textwidth]{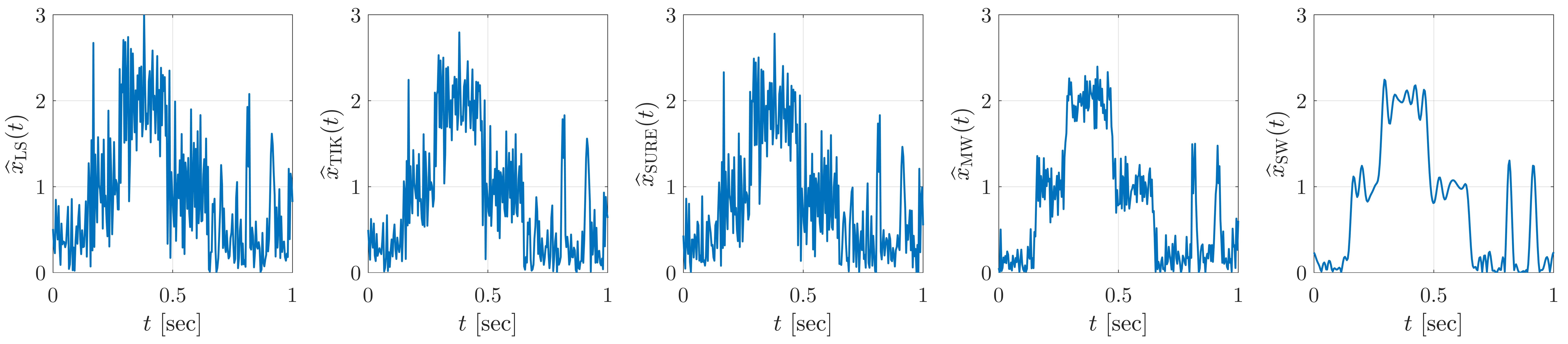}\vspace{-0.1cm}
			\caption{}
			\label{fig:typical_deconvolution_output_at_SNR_out_20dB}
		\end{subfigure}
	\begin{subfigure}{\textwidth}
		\includegraphics[width=\textwidth]{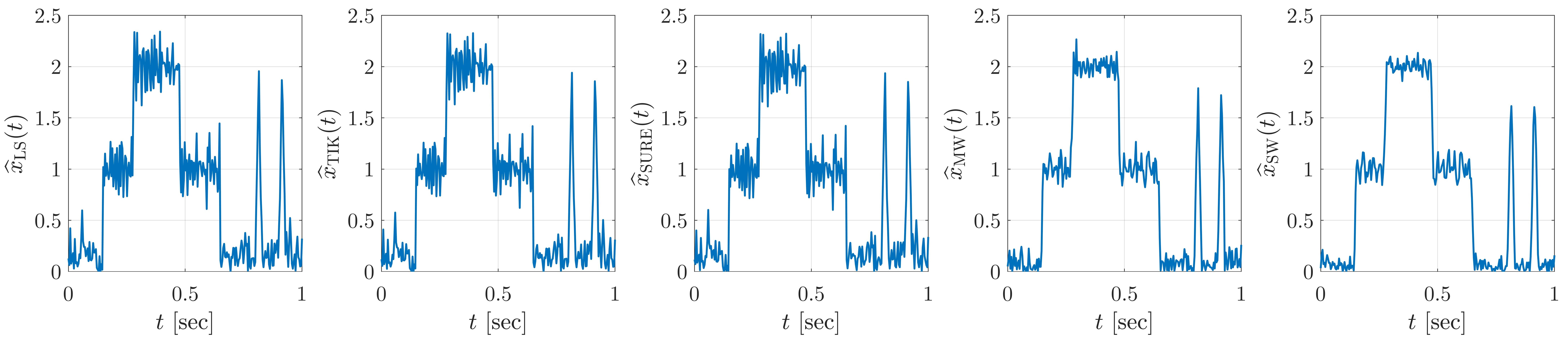}\vspace{-0.1cm}
		\caption{}
		\label{fig:typical_deconvolution_output_at_SNR_out_30dB}
	\end{subfigure}
		\caption{\addraRI{Typical time-domain outputs of the various deconvolution methods (the estimated LIDAR signals) at an output SNR level of (a) $20$ dB (b) $30$ dB. As in Fig.\ \ref{fig:MSE_vs_SNR_out_LIDAR}, for MW we set $q=2$. It is observed that the most accurate estimate is produced by our proposed ``Self-Wiener" estimator (most right).}}
		\label{fig:typical_deconvolution}
	\end{figure*}
}

Finally, \delra{we present in }Fig.\ \ref{fig:nonGaussiannoise}\addra{ presents} the \delra{analytically predicted}\addra{formula} \eqref{predictedMSEofSW} and \addra{the }empirical MSEs of the SW estimator for the signals \eqref{signalsimulation1}--\eqref{signalsimulation3}, however now for time-domain measurements $y[n]$ contaminated either by Laplace or by uniform distributed noise $v[n]$. As expected, an excellent fit is evident---due to the DFT \eqref{DFTdef}, \delra{and }by virtue of the CLT \cite{heyde1974central}, the frequency-domain noise $V[k]$ is approximately distributed as CN. This is in compliance with our assumption on the noise's CN distribution {\myfontb\emph{per frequency}} in Section \ref{sec:MSEanalysis}.\addraRI{

We note that when additional prior knowledge of the time-domain noise distribution and/or the input signal structure is available, then other potentially more accurate estimators of $\sigma_v$ are possible. For example, when the time-domain noise $v[n]$ is known a priori to be independent, identically distributed with a Laplace or uniform distribution, and the input signal can be assumed to be largely smooth, $\sigma_v$ can be estimated directly in the time-domain; For details, see \cite{john2021adaptive}, Subsection III-F.
}
\vspace{-0.2cm}
\addra{
\subsection{Reconstruction of a Blurred LIDAR Signal}\label{subsec:LIDARsignal}
We now compare all five methods under consideration in the following benchmark scenario. The unknown continuous-time $x(t)$ is the ideal LIDAR signal depicted in Fig.\ \ref{fig:LIDAR} (cf.\ Figure 1 in \cite{johnstone2004wavelet} and Figure 1a in \cite{cavalier2007wavelet}). The received output signal $y[n]$ is then \eqref{modeleq}---the sampled input $x[n]\triangleq x(n\Delta T)$, with $\Delta T\triangleq\frac{1}{N}$, convolved with a system $h[n]$, whose associated frequency response $H[k]$ is given by \eqref{Frequencyresponse} with $\alpha=0.75$, contaminated by additive white Gaussian noise. We set $N=256=2^8$, which is similar to the block size used in \cite{wu2011comparison}, and is also is FFT-compatible. The log-magnitude of this frequency response $H[k]$, also termed ``smooth blur" due to its low pass filter nature, is depicted in Fig.\ \ref{fig:smoothblur} and is approximately equivalent to one being used in \cite{johnstone2004wavelet}, cf.\ Figure 2a therein.

Figure \ref{fig:MSE_vs_SNR_out_LIDAR} presents the MSE \eqref{costfunctionFreq} vs.\ the average output SNR \eqref{averageoutputSNR}. As seen, none of the methods above uniformly dominates all others. However, above an average output SNR of $-10$[dB], the SW estimator yields the lowest MSE of all considered methods, some by a large margin.\addraRI{ Figures \ref{fig:typical_deconvolution_output_at_SNR_out_20dB} and \ref{fig:typical_deconvolution_output_at_SNR_out_30dB} present typical time-domain outputs of the various deconvolution methods (i.e., the estimated LIDAR signals) at output SNR levels of $20$ dB and $30$ dB, respectively. The enhanced accuracy attained by the SW estimator is evident.}\comment{ From a certain, sufficiently high SNR level (e.g., $\text{SNR}_{\text{out,avg}}=30$[dB]), the differences in the MSE of these methods are rather negligible.}

This simulation demonstrates that while the SW estimator does not dominate all other methods, it is quite performant even for signals which are not necessarily frequency-domain sparse or bandlimited. Indeed, our proposed solution, which is shown analytically to provide asymptotically optimal performance in the high SNR regime (Theorem \ref{theorem4}) and noise suppression in the low SNR regime (Theorem \ref{theorem3}), \emph{does not} require any application specific tuning, and exhibits robustness for a variety of input signals and noise distributions.
}\vspace{-0.2cm}
\section{Conclusion}\label{sec:conclusion}
In the framework of deterministic signals reconstruction, we presented a non-iterative, fully data-driven robust deconvolution method, which is based on mild assumptions regarding the unknown signal and the noise. Our method is termed ``Self-Wiener" as it tries to mimic the optimal Wiener-like deconvolution filter, but uses its own output to estimate the unknown SNR of the signal. We presented an analytical performance analysis of the proposed estimator, which enables to accurately assess its predicted performance, and thus to better characterize its strengths and weaknesses. The performance gain over other common (not necessarily) data-driven alternatives was illustrated in simulations, reaching up to almost an order of magnitude reduction in the \delra{residual }MSE relative to these methods.
\addra{The asymptotic optimality of the SW estimator suggests that it may be extended to other problems, such as signal denoising. Potential directions for future research include application of the SW estimator---or a modified/extended version thereof---in such cases, possibly incorporating some (specific domain) prior knowledge on the input signal.}
\appendices

\section{Proof of Theorem \ref{theorem3}}\label{AppALowSNR}
To show that in the low SNR regime the MSE \eqref{MSE_of_SW} can be approximated by \delra{the expression }\eqref{MSE_of_SW_low_SNR_theorem3}, we expand the quantities \eqref{finalexpprob}, \eqref{condexpc1} and \eqref{condexpc2} at $|\eta|\ll1$.
Starting with \eqref{finalexpprob}, we have by definition
\begin{equation}\label{derivationofapproxp}
p=\Pr\left(|Z|>2\right)=e^{-|\eta|^2}\int_{2\sqrt{2}}^{\infty}{xe^{-\frac{x^2}{2}}I_{0}(\sqrt{2}|\eta|x)\text{d}x},
\end{equation}
where $I_0$ is the modified Bessel function of the first kind. Since $p$ is analytic in $|\eta|$, performing a Taylor expansion of \eqref{derivationofapproxp} and using known results regarding the Bessel function yields\addra{ \cite{abramowitz1948handbook}}
\begin{equation}\label{resultofapproxp}
p=\int_{2\sqrt{2}}^{\infty}{xe^{-\frac{x^2}{2}}\text{d}x}+\mathcal{O}\left(|\eta|^2\right)=e^{-4}+\mathcal{O}\left(|\eta|^2\right).
\end{equation}

Next, we turn to \eqref{condexpc1}, and write it as
\begin{equation*}\label{Taylorofcondexpc1}
\Eset\left[\left.\widehat{X}_{\sw}\right|\left|Z\right|>2\right]\triangleq g_1(\eta)=g_1(0)+\mathcal{O}\left(|\eta|\right).
\end{equation*}
Recall that $\left.Z\right|_{\eta=0}=\widetilde{V}$, thus
\begin{equation*}\label{lowsnrapproxzeroexpectation}
g_1(0)=\frac{\sqrt{S_v}}{H}\Eset\left[\left.\tfrac{2\left(\widetilde{V}^*\right)^{-1}}{1-\sqrt{1-4\left|\widetilde{V}\right|^{-2}}}\right|\left|\widetilde{V}\right|>2\right].
\end{equation*}
Since $\widetilde{V}\sim\mathcal{CN}(0,1)$, it is invariant to rotations. As the domain $\left|\widetilde{V}\right|>2$ is \delra{also }circularly symmetric, \addra{we have}\delra{it follows that} $g_1(0)=0$. Hence,
\begin{equation}\label{Taylorofcondexpc1result}
\Eset\left[\left.\widehat{X}_{\sw}\right|\left|Z\right|>2\right]=\mathcal{O}\left(|\eta|\right).
\end{equation}
Similarly, we write \eqref{condexpc2} as
\begin{equation}\label{Taylorofcondexpc2}
\Eset\left[\left.\left|\widehat{X}_{\sw}\right|^2\right|\left|Z\right|>2\right]\triangleq g_2(\eta)=g_2(0)+\mathcal{O}\left(|\eta|\right).
\end{equation}
Using \eqref{sw_shrinkage_form}, \eqref{effnoise} and the relation $\widehat{X}_{\LS}=\tfrac{\sqrt{S_v}}{H}Z$, we have
\begin{equation*}
\begin{gathered}\label{lowsnrapproxsecondexpectation}
g_2(0)=\Eset\left[\left.\left|\widehat{X}_{\sw}\right|^2\right|\left|\widetilde{V}\right|>2\right]=\\
\frac{S_v}{4|H|^2}\cdot\Eset\left[\left.\left|\widetilde{V}\right|^2\left(1+\sqrt{1-4|\widetilde{V}|^{-2}}\,\right)^2\right|\left|\widetilde{V}\right|>2\right]=\\
\frac{\sigma_{\text{eff}}^2}{2}\cdot\Eset\left[\left.\left|\widetilde{V}\right|^2+\sqrt{\left|\widetilde{V}\right|^4-4\left|\widetilde{V}\right|^2}-2\right|\left|\widetilde{V}\right|>2\right].
\end{gathered}
\end{equation*}
Here, $\left|\widetilde{V}\right|^2\triangleq\xi^2/2$, where $\xi^2$ follows a chi-square distribution with two degrees of freedom. Its density is simply an exponential with rate $1/2$, namely $f(t)=\frac{1}2\exp(-t/2)$. Furthermore, the domain of integration is $\xi^2 >8$, and as we showed in \eqref{resultofapproxp}, $\left.p\right|_{\eta=0}=\left.\Pr\left(|Z|>2\right)\right|_{\eta=0}=\Pr\left(\xi^2>8\right)=e^{-4}$. Hence,
\begin{align*}
&\frac{\sigma_{\text{eff}}^2}{2}\cdot\Eset\left[\left.\left|\widetilde{V}\right|^2+\sqrt{\left|\widetilde{V}\right|^4-4\left|\widetilde{V}\right|^2}-2\right|\left|\widetilde{V}\right|>2\right]=\nonumber\\
&\frac{\sigma_{\text{eff}}^2}{2}\cdot \frac{1}{\left.p\right|_{\eta=0}} \cdot\int_8^\infty \left[\frac{t}2+\sqrt{\frac{t^2}{4}-2t}-2\right]\frac{1}{2}e^{-\frac{t}{2}} \text{d}t=\sigma_{\text{eff}}^2\cdot\frac{\rho}{e^{-4}},
\end{align*}
where the scalar $\rho$ is given by
\begin{equation}\label{definitionofrho}
\begin{aligned}
\rho &=\frac12 \int_8^\infty \left[\frac{t}2+\sqrt{t^2/4-2t}-2\right]\frac{1}{2} e^{-\frac{t}{2}} \text{d}t\\
&=\frac{1}2\left[5e^{-4}+2e^{-2}K_1(2) - 2e^{-4}\right]\approx 0.0464,
\end{aligned}
\end{equation}
and $K_1$ is the modified Bessel function of the second kind. \delra{We thus conclude that}\addra{Hence,} $g_2(0)=\sigma_{\text{eff}}^2\cdot\rho\cdot e^4$. Therefore, \eqref{Taylorofcondexpc2} reads
\begin{equation}\label{Taylorofcondexpc2result}
\Eset\left[\left.\left|\widehat{X}_{\sw}\right|^2\right|\left|Z\right|>2\right]=\frac{\sigma_{\text{eff}}^2\cdot\rho}{e^{-4}}+\mathcal{O}\left(|\eta|\right).
\end{equation}
\delra{Finally, s}\addra{S}ubstituting \eqref{resultofapproxp}, \eqref{Taylorofcondexpc1result} and \eqref{Taylorofcondexpc2result} into \eqref{MSE_of_SW} gives \eqref{MSE_of_SW_low_SNR_theorem3}\delra{, as required}. \hfill $\blacksquare$

\section{Proof of Theorem \ref{theorem4}}\label{AppAHighSNR}
To prove the theorem, we shall use the following lemma.
\begin{lem}\label{lemma1}
Let $Z=\eta+\widetilde{V}$, where $\widetilde{V}\sim\mathcal{CN}(0,1)$. Then, as $|\eta|\to\infty$,
\begin{align}
&\Eset\left[\left.\frac{1}{|Z|^2}\right||Z|>2\right]=\frac{1}{|\eta|^2}+\mathcal{O}\left(\frac{1}{|\eta|^4}\right),\label{lemmaeq1}\\
&\Eset\left[\left.\frac{\widetilde{V}}{\eta|Z|^2}\right||Z|>2\right]=\mathcal{O}\left(\frac{1}{|\eta|^4}\right).\label{lemmaeq2}
\end{align}
\end{lem}
\noindent\textit{Proof of Lemma \ref{lemma1}:} Let us write
\begin{equation*}\label{zintermsofeta}
\frac{1}{|Z|^2}=\frac{1}{|\eta+\widetilde{V}|^2}=\frac{1}{|\eta|^2}\cdot\frac{1}{1+2\Re\left\{\widetilde{V}/\eta\right\}+|\widetilde{V}|^2/|\eta|^2}.
\end{equation*}
As $|\eta|\to\infty$, with high probability, up to exponentially small terms in $|\eta|$, $|\widetilde{V}/\eta|\ll 1$. Under this event, we may thus perform a Taylor expansion $\frac{1}{1-x}=\sum_{n=0}^{\infty}x^n$
to obtain
\begin{equation}\label{zintermsofetaTaylor}
\begin{aligned}
\frac{1}{|Z|^2}=\frac{1}{|\eta|^2}\cdot\bigg(1-2\Re\left\{\widetilde{V}/\eta\right\}\bigg)+\mathcal{O}_P\left(\frac{1}{|\eta|^4}\right).
\end{aligned}
\end{equation}
We now take the conditional expectation. However, since $|\eta|\gg 1$, we may neglect the condition $|Z|>2$, and perform the integration over all of the domain of $\widetilde{V}$. This introduces a negligible error, exponentially small in $|\eta|$. 
Since $\Eset\left[\widetilde{V}\right]=0$, taking the expectation of \eqref{zintermsofetaTaylor} gives \eqref{lemmaeq1}.

To prove \eqref{lemmaeq2}, we again use \eqref{zintermsofetaTaylor} to have
\begin{equation}\label{secondargumentoflemma}
\frac{\widetilde{V}}{\eta|Z|^2}=\frac{1}{|\eta|^2}\cdot\bigg(\frac{\widetilde{V}}{\eta}-2\frac{\widetilde{V}}{\eta}\Re\left\{\frac{\widetilde{V}}{\eta}\right\}\bigg)+\mathcal{O}_P\left(\frac{1}{|\eta|^4}\right).
\end{equation}
We now take the expectation\delra{ in the same way} as\addra{ done} above, namely neglect the condition $|Z|>2$ and integrate over all of the domain of $\widetilde{V}$, thus introducing an error exponentially small in $|\eta|$. Since
\begin{equation*}
\Eset\left[\left(\widetilde{V}/\eta\right)\Re\left\{\widetilde{V}/\eta\right\}\right]=\frac{1}{2}\cdot\Eset\left[\left|\widetilde{V}/\eta\right|^2\right]=\frac{1}{2|\eta|^2},
\end{equation*}
taking the expectation of \eqref{secondargumentoflemma} gives \eqref{lemmaeq2}  \hfill $\blacksquare$

\noindent\textit{Proof of Theorem \ref{theorem4}:} Note that for $|\eta|\gg 1$, with high probability (up to deviations exponentially small in $|\eta|$), 
$|Z|=\left|\eta + \widetilde{V}\right|\gg 1$. Hence, using the Taylor expansion 
$\sqrt{1-\epsilon} = 1-\frac{\epsilon}2 + \mathcal{O}(\epsilon^2)$ in \eqref{sw_shrinkage_form} gives
\begin{equation}\label{sw_est_Taylor_expansion}
\widehat{X}_{\sw}=\left(X+\frac{\sqrt{S_v}}{H}\widetilde{V}\right)\cdot\left(1-|Z|^{-2}+\mathcal{O}_P\left(|\eta|^{-4}\right)\right).
\end{equation}

Let us start with \eqref{condexpc1}. Using \eqref{sw_est_Taylor_expansion}, we have that
\begin{align}
&\Eset\left[\left.\widehat{X}_{\sw}\right|\left|Z\right|>2\right]=\nonumber\\
&X\cdot\Eset\left[\left.1-|Z|^{-2}+\mathcal{O}_P\left(|\eta|^{-4}\right)\right||Z|>2\right]+\label{firsttermincondexpc1}\\
&\frac{\sqrt{S_v}}{H}\cdot\Eset\left[\left.\widetilde{V}\left(1-|Z|^{-2}+\mathcal{O}_P\left(|\eta|^{-4}\right)\right)\right||Z|>2\right].\label{secondtermincondexpc1}
\end{align}
Regarding the term \eqref{firsttermincondexpc1}, using \eqref{lemmaeq1} of Lemma \ref{lemma1},
\begin{align*}
&X\cdot\Eset\left[\left.1-|Z|^{-2}+\mathcal{O}_P\left(|\eta|^{-4}\right)\right||Z|>2\right]=\\
&X\cdot\Bigg(1-\frac{1}{|\eta|^2}+\mathcal{O}\left(\frac{1}{|\eta|^4}\right)\Bigg).
\end{align*}
As for \eqref{secondtermincondexpc1}, up to exponentially small terms in $|\eta|$, $\Eset\left[\left.\widetilde{V}\right||Z|>2\right]=0$. Therefore, using \eqref{lemmaeq2} of Lemma \ref{lemma1},
\begin{align*}
&\frac{\sqrt{S_v}}{H}\cdot\Eset\left[\left.\widetilde{V}\left(1-|Z|^{-2}+\mathcal{O}_P\left(|\eta|^{-4}\right)\right)\right||Z|>2\right]=\\
&X\cdot\left(\Eset\left[\left.\frac{\widetilde{V}}{\eta|Z|^2}\right||Z|>2\right]+\mathcal{O}\left(\frac{1}{|\eta|^4}\right)\right)=X\cdot\mathcal{O}\left(\frac{1}{|\eta|^4}\right)
\end{align*}
Hence, the above gives
\begin{equation}\label{approx_of_condexpc1}
\Eset\left[\left.\widehat{X}_{\sw}\right|\left|Z\right|>2\right]=X\cdot\Bigg(1-\frac{1}{|\eta|^2}+\mathcal{O}\left(\frac{1}{|\eta|^4}\right)\Bigg).
\end{equation}

Next, we turn to analyze \eqref{condexpc2}. Note that
\begin{equation*}
\Big[1-|Z|^{-2}+\mathcal{O}_P\left(|\eta|^{-4}\right)\Big]^2=1-2|Z|^{-2}+\mathcal{O}_P\left(|\eta|^{-4}\right).
\end{equation*}
Thus, using the relation $\left|\widehat{X}_{\LS}\right|^2=\sigma_{\text{eff}}^2\cdot|Z|^2=\tfrac{|X|^2}{|\eta|^2}\cdot|Z|^2$,
\begin{equation*}
\begin{aligned}
\left|\widehat{X}_{\sw}\right|^2&=\frac{|X|^2}{|\eta|^2}\cdot|Z|^2\cdot\Big(1-2|Z|^{-2}+\mathcal{O}_P\left(|\eta|^{-4}\right)\Big)\\
&=|X|^2\cdot\left(\frac{|Z|^2}{|\eta|^2}-\frac{2}{|\eta|^2}+\mathcal{O}_P\left(|\eta|^{-4}\right)\right).
\end{aligned}
\end{equation*}
Using $\Eset\left[|Z|^2\right]=|\eta|^2+1$ and similar arguments as before yield that
\begin{equation}\label{approx_of_condexpc2}
\Eset\left[\left|\widehat{X}_{\sw}\right|^2\left.\right||Z|>2\right]=|X|^2\cdot\Bigg(1-\frac{1}{|\eta|^2}+\mathcal{O}\left(\frac{1}{|\eta|^4}\right)\Bigg).
\end{equation}
Since $|X|^2/|\eta|^4=\sigma_{\text{eff}}^2/\text{SNR}_{\text{out}}$, substituting \eqref{approx_of_condexpc1} and \eqref{approx_of_condexpc2} into \eqref{MSE_of_SW} gives \eqref{MSE_of_SW_high_SNR_theorem4}. \hfill $\blacksquare$

\section{MSE Analysis of Real-Valued DFT Components}\label{AppArealDFT}
The analysis for the real-valued DFT bins, corresponding to the indices $k=0,N/2$, is very similar to the analysis presented in Section \ref{sec:MSEanalysis} and above. In these cases $\widetilde{V}\sim\mathcal{N}(0,1)$, thus the probability $p[k]$ reads
\begin{equation}\label{pkforrealvaluedcase}
k=0,\tfrac{N}{2}:p[k]=\Pr(|Z[k]|>2)=Q(2-\eta[k])+Q(2+\eta[k]),
\end{equation}
where $Q(\cdot)$ is the Q-function: $Q(x)\triangleq\tfrac{1}{2\pi}\int_{x}^{\infty}e^{-0.5t^2}\text{d}t$ (recall that $\eta[k]$ is real-valued for $k=0,N/2$). 

For the low SNR approximation, it is easy to verify that,
\begin{equation}\label{papproxforrealvaluedcase}
|\eta|\ll1: p=2Q(2)+\mathcal{O}(|\eta|),
\end{equation}
and that \eqref{condexpc1} is still
\begin{equation*}
\Eset\left[\left.\widehat{X}_{\sw}\right|\left|Z\right|>2\right]=\mathcal{O}\left(|\eta|\right),
\end{equation*}
from symmetry considerations. However, \eqref{condexpc2} now reads
\begin{equation}\label{approx5forrealvaluedcase}
\Eset\left[\left.\left|\widehat{X}_{\sw}\right|^2\right|\left|Z\right|>2\right]=\frac{|X|^2\varrho}{2Q(2)|\eta|^2}+\mathcal{O}\left(|\eta|\right),
\end{equation}
where
\begin{align*}\label{defofvarrhoforrealvaluedcase}
\varrho&\triangleq\int_{2}^{\infty}\left[v^2-2+\sqrt{v^4-4v^2}\,\right]\frac{1}{\sqrt{2\pi}}e^{-\frac{v^2}{2}}\text{d}v\\
&=\frac{e^{-2}}{2}+e^{-2}\sqrt{\frac{2}{\pi}}-Q(2)\approx0.1529,
\end{align*}
an order of magnitude greater than $\rho$ in \eqref{definitionofrho} of the complex-valued case. Using the updated terms \eqref{papproxforrealvaluedcase} and \eqref{approx5forrealvaluedcase}, the MSE \eqref{MSE_of_SW} at frequencies with low SNR, namely $|\eta|\ll1$, for the \emph{real-valued} DFT components is given by
\begin{equation*}\label{approx_MSE_of_SW_low_SNR_real_valued}
\begin{aligned}
\text{MSE}_{\sw}&=|X|^2+\varrho\cdot\sigma_{\text{eff}}^2+\mathcal{O}\left(|\eta|\right)\\
&=\left|X\right|^2\cdot\left(1+\frac{\varrho}{\text{SNR}_{\text{out}}}\right)+\mathcal{O}\left(\sqrt{\text{SNR}_{\text{out}}}\right),
\end{aligned}
\end{equation*}
so the MSE is greater for the real-valued DFT components.

The high SNR approximation is also obtained in the same fashion, only now $Z\sim\mathcal{N}(\eta,1)$. It follows that for $|\eta|\gg1$, \eqref{lemmaeq1} and \eqref{lemmaeq2} from Lemma \ref{lemma1} hold in this case as well. Accordingly, using the fact that $\Eset[Z^2]=\eta^2+1$ and similar arguments as in Theorem \ref{theorem4}, it is easy to verify that \eqref{approx_of_condexpc1} and \eqref{approx_of_condexpc2} hold true when $Z$ is normal, rather than CN. Therefore, the MSE \eqref{MSE_of_SW} at frequencies with high SNR, namely $|\eta|\gg1$, for the \emph{real-valued} DFT components is given by 
\begin{equation*}\label{MSE_of_SW_high_SNR_real_valued}
\begin{aligned}
\text{MSE}_{\sw}&=(1-p)\cdot\left|X\right|^2+p\cdot\frac{|X|^2}{|\eta|^2}+\mathcal{O}\left(\frac{|X|^2}{|\eta|^4}\right)\\
&=(1-p)\cdot\left|X\right|^2+p\cdot\sigma_{\text{eff}}^2+\mathcal{O}\left(\frac{\sigma_{\text{eff}}^2}{\text{SNR}_{\text{out}}}\right),
\end{aligned}
\end{equation*}
with $p=Q(2-\sqrt{\text{SNR}_{\text{out}}})+Q(2+\sqrt{\text{SNR}_{\text{out}}})$ as in \eqref{pkforrealvaluedcase}.

\addra{\section{Intermediate SNR Approximation of MSE}\label{AppAintermediateSNR}
In this Appendix we provide the derivation leading to the approximated MSE expression \eqref{predictedMSEofSW}. \addraRI{First, note that from \eqref{MSElawoftotalexpectation}, the MSE can be viewed as a convex combination of a ``high-SNR" term and a ``low-SNR" term, where we recall $p\xrightarrow[]{\text{SNR}_{\text{out}}\to\infty}1$ and $p\xrightarrow[]{\text{SNR}_{\text{out}}\to0}e^{-4}\ll1$. Therefore, we choose a value $\tau>0$ such that}\delra{Assume that} the low and high SNR approximations\addraRI{ \eqref{MSE_of_SW_low_SNR_theorem3} and \eqref{MSE_of_SW_high_SNR_theorem4}} are sufficiently accurate for $\text{SNR}_{\text{out}}<-\tau$[dB] and $\text{SNR}_{\text{out}}>\tau$[dB], respectively\delra{, for some $\tau\in\Rset^+$}. In the intermediate SNR interval $[-\tau,\tau]$[dB], \addraRI{as a heuristic solution, }we propose to approximate the predicted SNR via the following interpolation. Recall that $\varepsilon^2_{\text{low}}$ and $\varepsilon^2_{\text{high}}$ denote the low \eqref{MSE_of_SW_low_SNR_theorem3} and high \eqref{MSE_of_SW_high_SNR_theorem4} MSE approximations, respectively. Now, compute the slope and intercept coefficients
\begin{equation}\label{slope_and_intercept}
a\triangleq \frac{\varepsilon^2_{\text{high}}-\varepsilon^2_{\text{low}}}{2\tau\text{[dB]}}, \quad b\triangleq \frac{\varepsilon^2_{\text{high}}+\varepsilon^2_{\text{low}}}{2},
\end{equation}
with which our approximation for the MSE in the intermediate SNR interval, namely for $\text{SNR}_{\text{out}}\in[-\tau,\tau]$[dB], is given by
\begin{equation}\label{AppAapproxMSEintermediateSNR}
\text{MSE}_{\sw}\underset{|\eta|^2<\tau}{\approx}a\cdot\text{SNR}_{\text{out}}\text{[dB]}+b.
\end{equation}
From our thorough empirical examinations, choosing $\tau=6$[dB] gives fairly accurate results, as evident from Fig.\ \ref{fig:predicted_MSE}.\addraRI{ Note further that $\tau=6\text{ [dB]}\approx3.9811$ implies that the signal power is approximately four times larger than the noise power, and this is sufficient for the high SNR approximation in Theorem \ref{theorem4} to be very accurate. A similar justification holds for the low SNR case, where $\tau=-6\text{ [dB]}\approx0.2512$.}}

\bibliography{Bibfile}
\bibliographystyle{unsrt}

\end{document}